\documentclass[twocolumn,amsmath,amssymb,prl,superscriptaddress]{revtex4-2}
\usepackage{times}
\usepackage[pdftex]{graphicx}
\usepackage{dcolumn}
\usepackage{bm}
\usepackage{amsmath}
\usepackage{indentfirst}
\usepackage{float}
\usepackage[colorlinks,linkcolor=blue,anchorcolor=blue,citecolor=blue]{hyperref}
\usepackage[dvipsnames]{xcolor}
\usepackage{braket}
\usepackage{multirow}
\usepackage{changes}
\usepackage{pifont}
\usepackage{makecell}
\usepackage{amsthm,amssymb,bbm,newtxmath}
\usepackage{mathrsfs}
\usepackage[noend]{algpseudocode}
\usepackage{array}
\usepackage{booktabs}
\usepackage{multirow}
\usepackage{stackengine}
\usepackage{algorithmicx,algorithm}
\usepackage{CJK}
\usepackage{pdfpages}
\usepackage{pgffor}

\makeatletter
\AtBeginDocument{\let\LS@rot\@undefined}
\makeatother

\def\supplementfilename{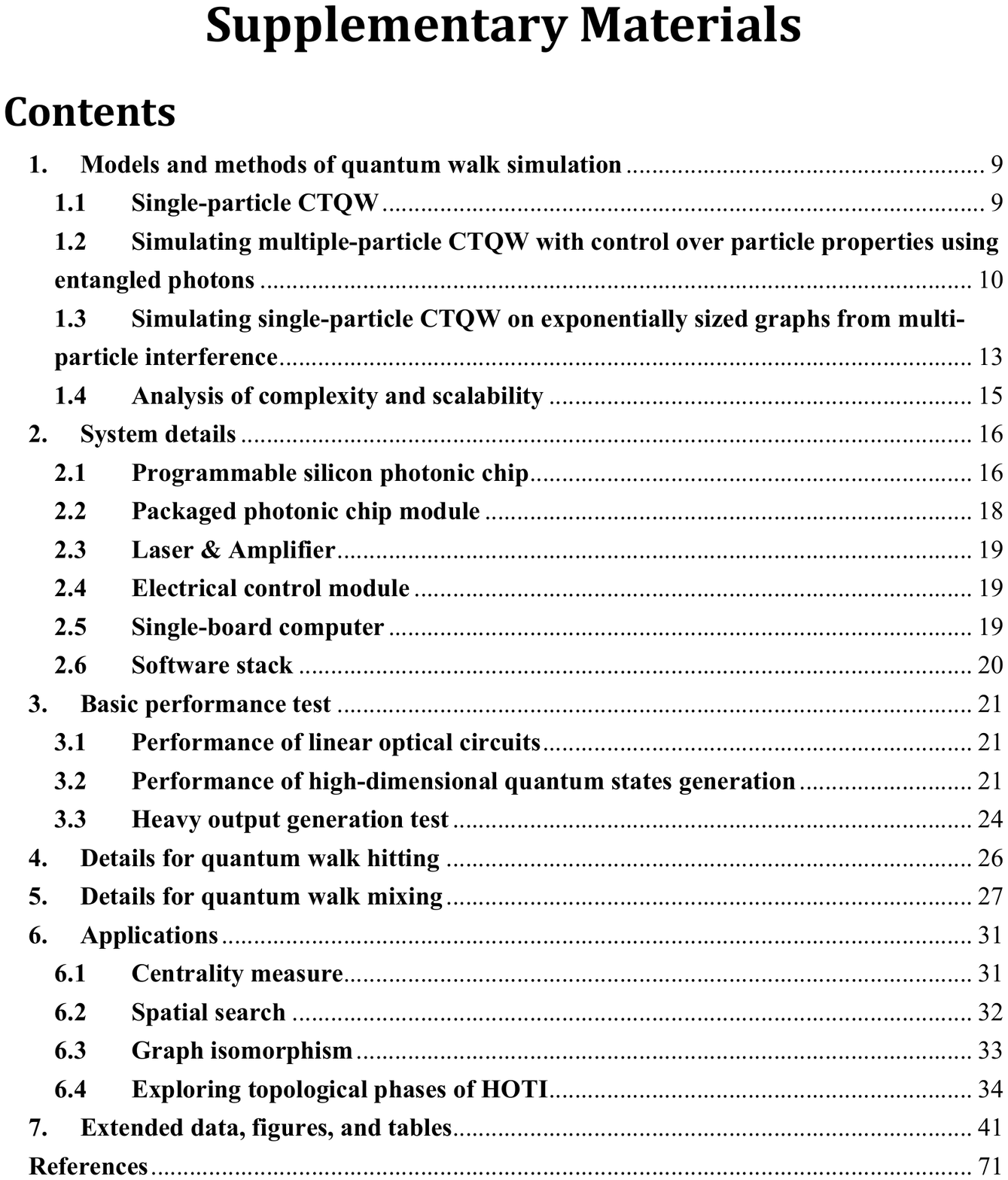}

\pdfximage{\supplementfilename}
\def\numbersupplementpages{\the\pdflastximagepages}

\newif\ifarXiv
\arXivtrue 

\hypersetup{colorlinks=true, citecolor=blue, urlcolor=blue, linkcolor=blue}

\begin{document}

\title{Large-scale full-programmable quantum walk and its applications}

\author{Yizhi Wang}
\thanks{These authors contributed equally to this work.}
\author{Yingwen Liu}
\thanks{These authors contributed equally to this work.}
\author{Junwei Zhan}
\author{Shichuan Xue}
\author{Yuzhen Zheng}
\author{Ru Zeng}
\author{\\Zhihao Wu}
\author{Zihao Wang}
\author{Qilin Zheng}
\author{Dongyang Wang}
\author{Weixu Shi}
\author{Xiang Fu}
\author{\\Ping Xu}
\affiliation{Institute for Quantum Information \& State Key Laboratory of High Performance Computing, College of Computer Science and Technology, National University of Defense Technology, Changsha 410073, China}
\author{Yang Wang}
\affiliation{National Innovation Institute of Defense Technology, AMS, Beijing 100071, China}
\author{Yong Liu}
\author{Jiangfang Ding}
\author{Guangyao Huang}
\affiliation{Institute for Quantum Information \& State Key Laboratory of High Performance Computing, College of Computer Science and Technology, National University of Defense Technology, Changsha 410073, China}
\author{Chunlin Yu}
\affiliation{China Greatwall Research Institute, China Greatwall Technology Group CO., LTD., Shenzhen 518057, China}
\author{Anqi Huang}
\affiliation{Institute for Quantum Information \& State Key Laboratory of High Performance Computing, College of Computer Science and Technology, National University of Defense Technology, Changsha 410073, China}
\author{Xiaogang Qiang}
\email{qiangxiaogang@gmail.com}
\affiliation{National Innovation Institute of Defense Technology, AMS, Beijing 100071, China}
\author{Mingtang Deng}
\author{Weixia Xu}
\author{Kai Lu}
\email{kailu@nudt.edu.cn}
\author{Xuejun Yang}
\author{Junjie Wu}
\email{junjiewu@nudt.edu.cn}
\affiliation{Institute for Quantum Information \& State Key Laboratory of High Performance Computing, College of Computer Science and Technology, National University of Defense Technology, Changsha 410073, China}


\begin{abstract}
    With photonics, the quantum computational advantage has been demonstrated on the task of boson sampling.
    Next, developing quantum-enhanced approaches for practical problems becomes one of the top priorities for photonic systems.
    Quantum walks are powerful kernels for developing new and useful quantum algorithms.
    Here we realize large-scale quantum walks using a fully programmable photonic quantum computing system.
    The system integrates a silicon quantum photonic chip, enabling the simulation of quantum walk dynamics on graphs with up to 400 vertices and possessing full programmability over quantum walk parameters, including the particle property, initial state, graph structure, and evolution time. 
    In the 400-dimensional Hilbert space, the average fidelity of random entangled quantum states after the whole on-chip circuit evolution reaches as high as 94.29$\pm$1.28$\%$.
    With the system, we demonstrated exponentially faster hitting and quadratically faster mixing performance of quantum walks over classical random walks, 
    achieving more than two orders of magnitude of enhancement in the experimental hitting efficiency and almost half of the reduction in the experimental evolution time for mixing.
    We utilize the system to implement a series of quantum applications, including measuring the centrality of scale-free networks, searching targets on Erd\"{o}s-R\'{e}nyi networks, distinguishing non-isomorphic graph pairs, and simulating the topological phase of higher-order topological insulators.
    Our work shows one feasible path for quantum photonics to address applications of practical interests in the near future.
\end{abstract}

\maketitle

Quantum computers have long been the anchor of hope for outperforming classical computers on a number of tasks \cite{Feynman1982,Shor1997}. 
Recently in bulk optics \cite{Zhong2020,Madsen2022}, quantum computational advantages have been demonstrated on a classically intractable problem, i.e., boson sampling. Besides, integrated photonics also showed the potential for implementing boson sampling \cite{Alexeev2021}. 
The standard boson sampling that constitutes multiple indistinguishable bosons undergoing coherent evolution in a Haar-random linear network can be viewed as an instance of quantum walks (QWs), while QWs define quantum dynamics of various particles on general graphs \cite{Aharonov1993}. 
QWs become promising quantum computation primitives since problem instances can be encoded into and further solved via the evolution dynamics of QWs with the properly chosen particle property and graph structure.
Many QW-based algorithms have shown quantum-enhanced performances in applications of practical interests, such as searching, analyzing, and learning complex networks \cite{Biamonte2019}.
QWs on designed networks can also model quantum dynamics in the fields of physics \cite{Qiang2016}, chemistry \cite{Bauer2020}, and biology \cite{Mohseni2008}, and even realize universal quantum computation \cite{Childs2013}.

\begin{figure*}[ht]
    \includegraphics[width=16.0cm]{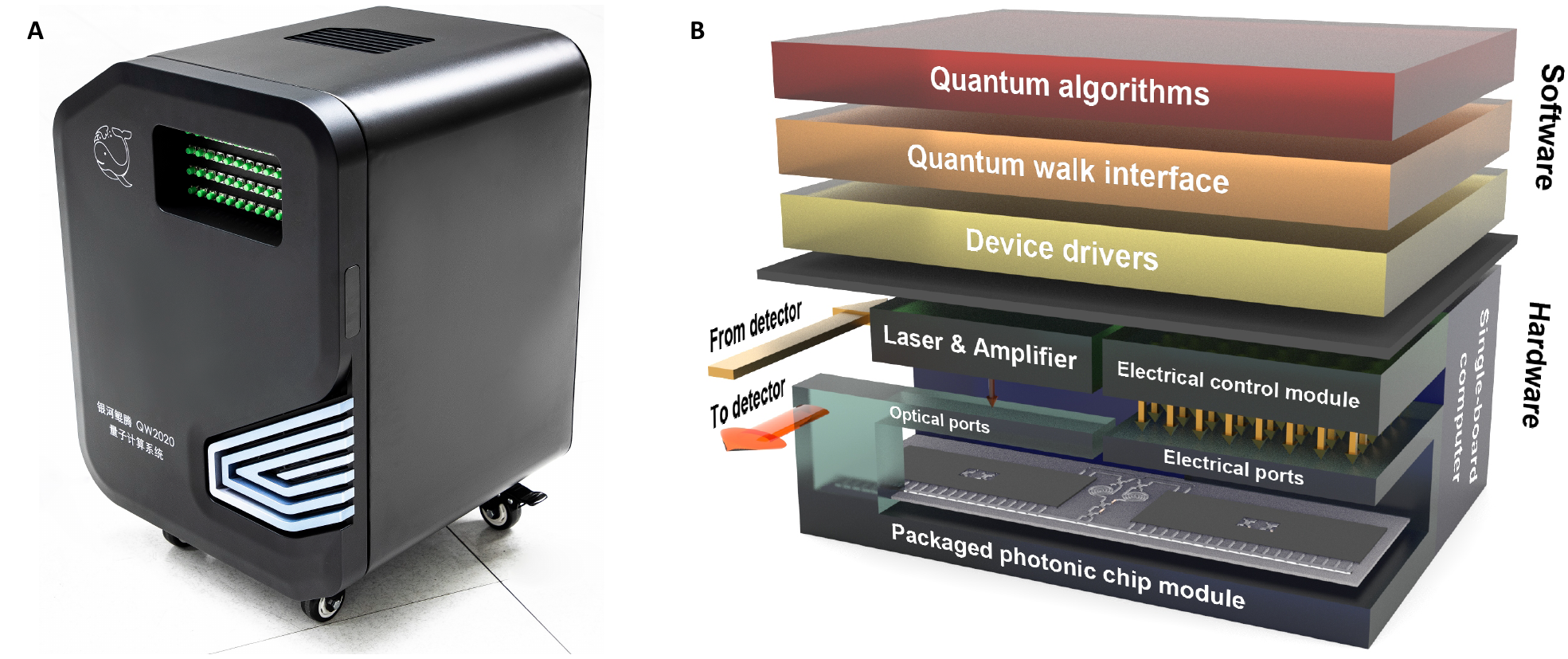}
    \caption{\label{fig:fig1}Overview of \emph{YH QUANTA QW2020} system. (A) The photograph of the whole system. The whole system is compactly contained within an 85cm$\times$60cm$\times$55cm portable case, excluding the detection module. (B) Schematic of the system stack. The software stack compiles quantum algorithms into quantum walk settings and then converts them to operations in hardware. At the bottom of the hardware stack is the packaged photonic chip module engineering two-photon states in a 400-dimensional Hilbert space, in which the embedded programmable silicon photonic chip is fully packaged. The control signals are transmitted via a self-developed 512-channel electronic control module. Pumped by an amplified laser and evolved through the chip, generated photon pairs are recorded by a peripheral detection module. The entire system stack is mastered by a small single-board computer. }
\end{figure*}

To exploit the potentials of QWs, physical apparatus with multiple particles and large-scale evolution graphs are essential. Moreover, the full programmability of particle properties and graph geometry is imperative to tackle different applications.
QWs have been experimentally demonstrated on various platforms, such as photons \cite{Harris2017,Benedetti2021,Xu2021,Qu2022}, superconducting qubits \cite{Gong2021}, trapped ions \cite{HuertaAlderete2020}, neutral atoms \cite{Karski2009}, and nuclear magnetic resonance systems \cite{Ryan2005}.
It remains challenging to simultaneously  combine all the capabilities for realizing the large-scale and full-programmable quantum walks.
In 2021, we reported a preliminary silicon photonic device  capable of simulating QW dynamics in a 25-dimensional Hilbert space with all parameter programmability \cite{Qiang2021}. 
Here, as shown in Fig.\ref{fig:fig1}A, we now present a full-stack photonic computing system, \emph{YH QUANTA QW2020} (\begin{CJK}{UTF8}{gkai}银河鲲腾\end{CJK}QW2020 in Chinese).
The system can implement full-programmable QWs in a 400-dimensional Hilbert space with improved accuracy and also take complete control over QW parameters, including the particle property, initial state, graph structure, and evolution time. 

Now consider the continuous-time QWs (CTQWs) of a single particle on an $N$-vertex graph $G$ with adjacency matrix $\boldsymbol{A}$. The system can be described by the Hamiltonian $H=\boldsymbol{A}$. The single-particle CTQW evolution follows 
\begin{equation}
  \ket{\psi(t)} = \mathrm{e}^{-iHt} \ket{\psi_{ini}},
\end{equation} 
where $H$ is the Hamiltonian on graph $G$, $\ket{\psi_{ini}}$ is the initial state and $\ket{\psi(t)}$ is the evolved state at time $t$.
When multiple particles get involved, the dimension of the corresponding Hilbert space can grow exponentially with the number of particles. 
A multiple-particle CTQW can be applied to simulate single-particle CTQW on an exponentially large graph, with the geometry of the large graph determined by the particle indistinguishability and exchange symmetry \cite{Qiang2021}. 
For example, the single-particle CTQW on an $N^P$-vertex Cartesian product graph $G^{(P)}_{D}$, of which the adjacency matrix is $\boldsymbol{A}^{(P)}_{D}=\boldsymbol{A}^{\oplus P}$, can be simulated by $P$ fully distinguishable particles evolution on graph $G$ \cite{Izaac2017}. When the particles are fully indistinguishable, the dimensions of the simulated larger graph for $P$ bosons (denoted as $G^{(P)}_{B}$) and $P$ fermions (denoted as $G^{(P)}_{F}$) are ${ N+P-1 \choose P }$ and ${ N \choose P }$, respectively \cite{Rudinger2012}. The explicit representation of the constructed large graph can be found in the supplementary materials \cite{SP}. 
These schemes make it possible that instead of directly implementing single-particle CTQW evolution on exponentially sized graph $G^{(P)}\in\{G^{(P)}_{D},G^{(P)}_{B},G^{(P)}_{F}\}$, we can simulate its dynamics via a multi-particle CTQW on a $N$-vertex graph $G$ with only polynomial resource cost \cite{SP}.

Experimentally, following our previous design \cite{Qiang2021}, we simulated two-particle CTQWs by sending each particle of the entangled two photons through identical copies of a single-particle CTQW evolution $U=\mathrm{e}^{-iHt}$.
The core of the system is one of the largest-scale programmable silicon quantum photonic chips. The chip monolithically integrates two spontaneous four-wave mixing photon-pair sources and two 20-mode universal linear optical circuits with fixed inputs. By tuning the generated entangled two-photon state, particles can be controlled with covering the entire spectrum from distinguishable to fully indistinguishable and from bosonic to fermionic exchange symmetry. Each of the two linear optical circuits can be configured to implement arbitrary state preparation and CTQW evolution in a Hilbert space with up to 20 dimensions. The whole chip enables the simulations of single-particle CTQWs on $G^{(2)}$ graphs with up to 400 vertices and has full control over all QW parameters, including particle property, initial state, graph structure, and evolution time. 

As shown in Fig.\ref{fig:fig1}B, the system is built following a hierarchical quantum computing stack \cite{Fu2017} that matches QW-based applications to physical hardware operations. 
At the top levels, quantum algorithms are expected to be modeled as QWs with various particle properties and evolution graphs. At the middle level,
the quantum walk interface translates algorithms to the corresponding parameters of QWs, and then compiles QW parameters into hardware operations. The device driver level interacts with the programmable photonic chip  to perform configuration and output detection of the chip. At the bottom of the stack, we have a packaged photonic chip module engineering two-photon states in a 400-dimensional Hilbert space, in which the embedded photonic chip is optically, electronically, and thermally packaged. A small single-board computer mastered all these distributed subsystems. 
We verify the high-precision of our system with one thousand randomized entangled quantum states in a 400-dimensional Hilbert space, which are generated by applying Haar-random unitary transformations to the programmable entangled two-photon. The average fidelity between the obtained and theoretical probability distributions of the evolved quantum states reaches $94.29\pm1.28\%$. Once assembled and calibrated, the system has been in continuous operation for over twenty months.

\begin{figure*}[ht]
    \includegraphics[width=16.0cm]{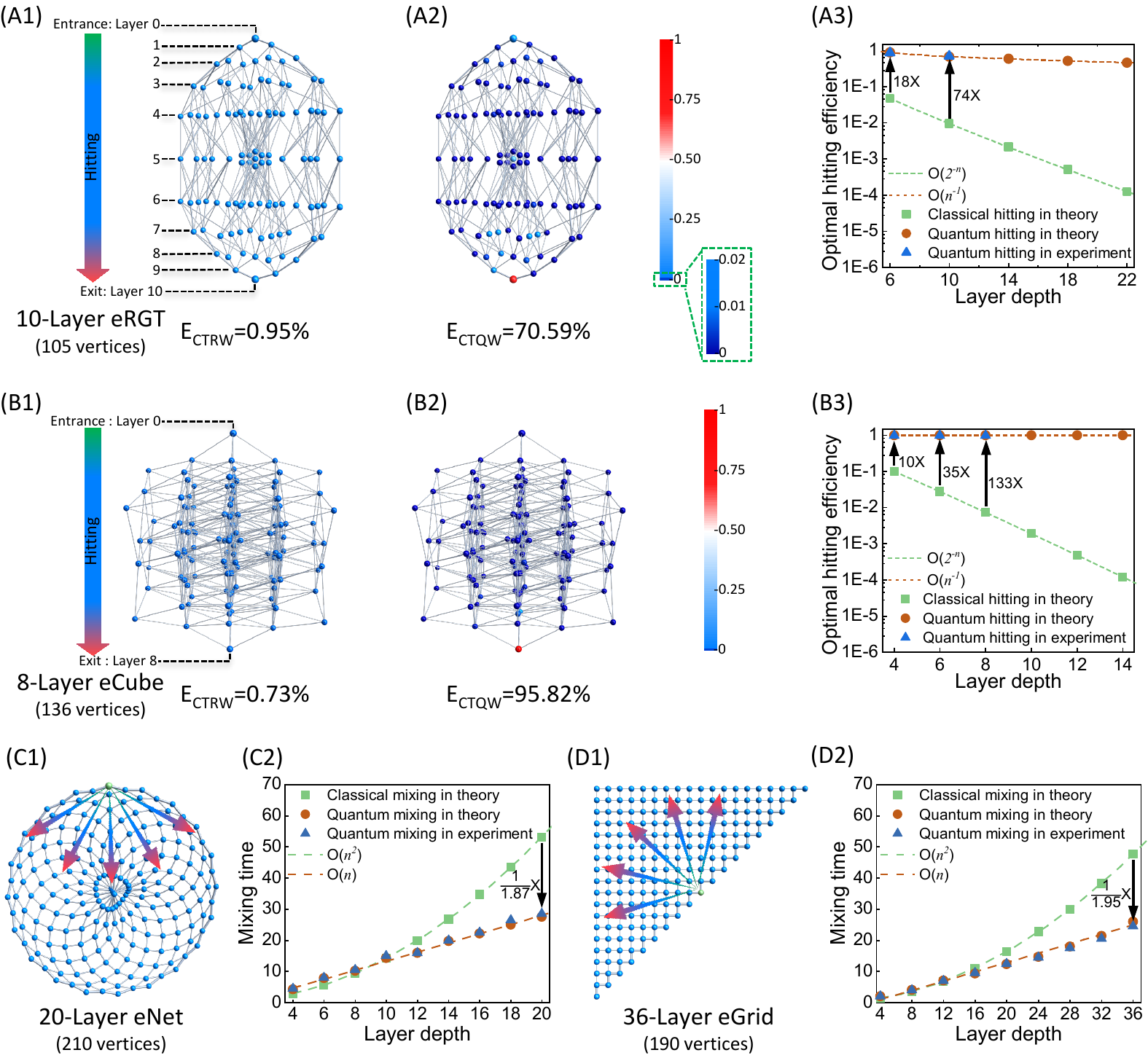}
    \caption{\label{fig:fig2}Exponentially fast hitting and quadratically fast mixing behaviors of quantum walks. (A) Exponentially fast hitting on RGTs. The 10-layer eRGT is generated via two-boson CTQW on 5-layer RGT. (A1) and (A2) compare the probability distributions when optimal hitting occurs for CTRW and CTQW. Colorbar showing the color scale is presented. In contrast with the optimal scenario (almost uniform distribution) in CTRW ($0.95\%$), quantum hitting efficiency ($70.59\%$) achieves more than two orders of enhancement. (A3) shows a fitted linear decrease trend for the quantum hitting efficiency, while classical hitting meets with an exponential drop. The experimentally obtained results (blue triangles) of CTQWs on 6-layer and 10-layer eRGTs are entirely consistent with theoretical predictions (brown circles). (B) Exponentially fast hitting on eCubes. Compared with the optimal hitting efficiency in CTRW (B1, $0.73\%$), CTQW (A2, $95.82\%$) also achieves more than two orders of enhancement on the 8-layer eCube generated via two-boson CTQW on the 4-layer hypercube. The exponential speedups of quantum hitting over classical hitting are shown in (A3) and (B3) with a logarithmic coordinate.  (C)-(D) Quadratically fast mixing on eNets and eGrids. (C1) and (D1) show the 20-layer eNet (generated via two-boson on 20-vertex cycle) and the 36-layer eGrid (generated via two-boson CTQW on 19-vertex line), with the start vertices colored in green. (C2) and (D2) compare the $\epsilon$-mixing evolution time ($\epsilon=0.25$) between quantum and classical mixing on eNets and eGrids with different sizes, respectively. The fitted data shows a clear linear trend for the quantum mixing, while the classical scenario needs a quadratically larger evolution time. }
\end{figure*}

\begin{figure*}[ht]
    \includegraphics[width=16.0cm]{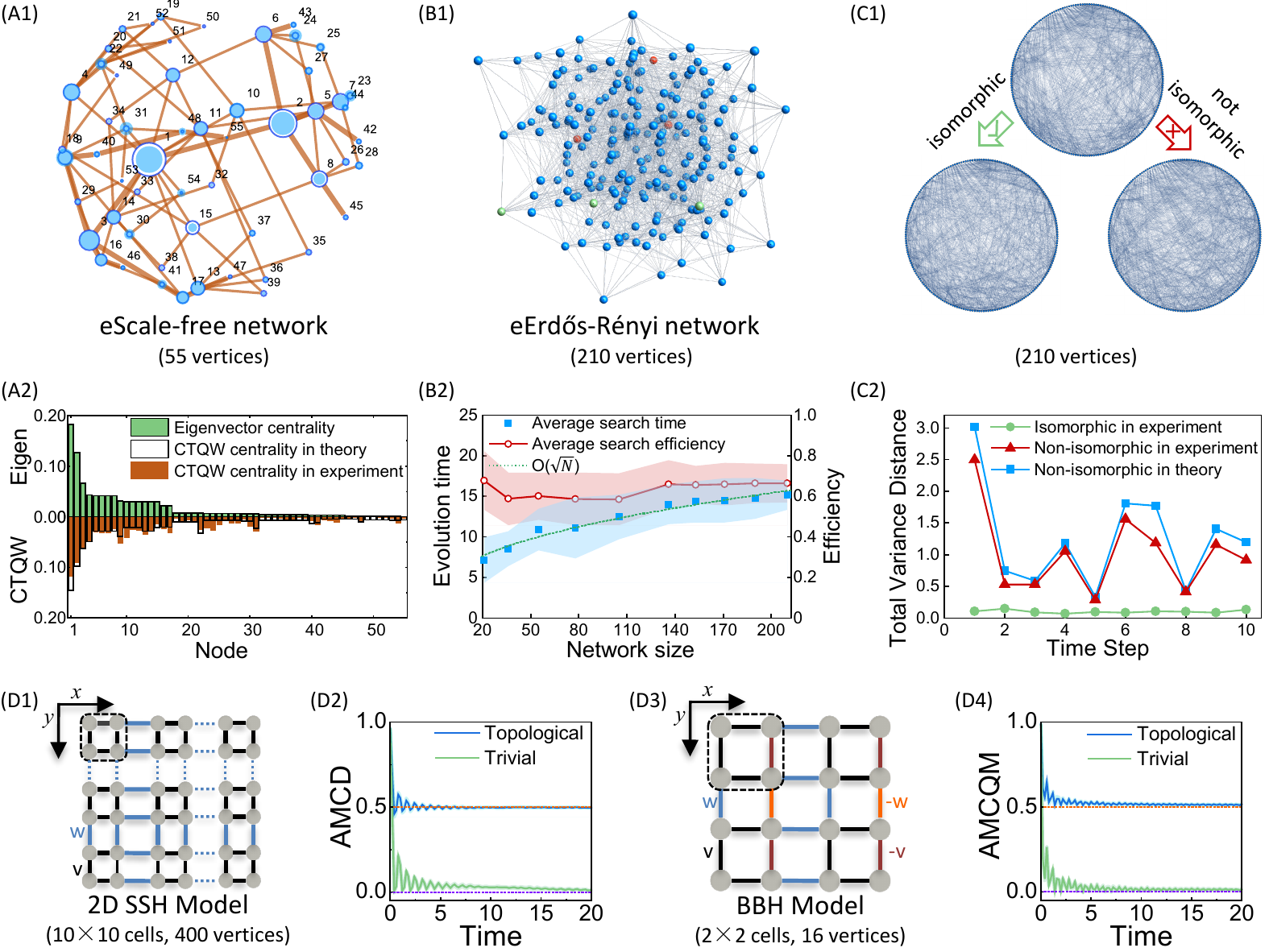}
    \caption{\label{fig:fig3}Demonstrations of CTQW-based applications using the system. (A) Centrality measure. (A1) A 55-vertex scale-free random network, with vertex size indicating centrality. The empty circles represent the theoretical eigenvector centrality, with experimentally determined CTQW centrality value overlaid. (A2) Comparison between eigenvector centralities and CTQW centrality. The similarity of eigenvector centrality and experimentally-obtained CTQW centrality is $95.68\%$. All centrality measures strongly agree on the top-ranked vertices, with slight variations for the lower-ranked vertices. (B) Spatial search test. (B1) An example of a 210-vertex Erdos-Renyi network. The initial positions of the quantum walker are colored in green, and the targets are brown. (B2) The statistical optimal search time grows as $\mathcal{O}(\sqrt{N})$ as network size $N$ scales from 10 to 210, while the search efficiency stabilizes around 0.5. We test spatial search on 100 Erdos-Renyi networks for each $N$. (C) Graph isomorphism determination. (C1) A 210-vertex graph (central), its isomorphic (left) and non-isomorphic (right) graphs. Graphs are plotted in a circular embedding layout. (C2) Total variance distance of CTQW-based graph certificates between the isomorphic pair remains stable around 0 (0.1013) during evolution, in contrast to the value of the non-isomorphic pair (1.2476 in theory and 1.0122 in the experiment) far greater than 0. The average of fidelities in the experiments for graph certificates reaches $94.76\pm1.08\%$. (D) Topological phase simulation. (D1) and (D3) present the 2D SSH model and the BBH model, respectively. Each unit cell (dashed box) consists of four vertices. The values of $v$ ($-v$) and $w$ ($-w$) represent the amplitudes of intracellular and intercellular hopping. (D2) shows that the long-time AMCDs on the y dimension of the 2D SSH model gradually approach the theoretical values. The experimental asymptotic results are $0.498\pm0.001$ for topological phases and $0.013\pm0.003$ for trivial phases. Similarly, in (D4) the long-time AMCDs values $0.513\pm0.002$ ($0.014\pm0.003$) for topological (trivial) phases also finally stabilize around the theoretical values of the BBH model.}
\end{figure*}

Before attempting to implement quantum algorithms based on QWs, it is first of profound importance to study their dynamics features. 
Once the underlying graphs of QWs increase to hundreds of vertices, quantum speedups could be obviously demonstrated in experiments. 
One of the most prominent QWs' features is the fast hitting ability on graphs, that is, to propagate from one vertex to another remote one more efficiently than classical random walks (CRWs) and even any known classical algorithms \cite{Childs2003}. 
On a hexagonal structure, quadratic speedups have been demonstrated \cite{Tang2018b}. However, the dynamics of exponentially fast hitting remain unexplored in experiments due to the need for complicated arrangements of exponentially increasing vertices.
With the capabilities to access the exponentially expanded Hilbert space, our system is able to carry out the first experimental observation of the hitting dynamics on eRGTs (that are extended-RGTs generated via two-boson CTQW on RGTs) with up to 105 vertices and eCube (extended-Cubes generated via two-boson CTQW on Hypercubes) with up to 136 vertices. 
In Fig.\ref{fig:fig2}A and Fig.\ref{fig:fig2}B, we compared the optimal hitting distribution and efficiency of CTQWs and CRWs. 
Nearly two orders of magnitude of enhancement in the hitting efficiency of CTQW experiments (0.7059, 0.9582) over CRWs (0.0095, 0.0073) are demonstrated on the 10-layer eRGT and 8-layer eCube, during which processes the average of the fidelities between the experimental and theoretical probability distributions are $96.28\pm1.86\%$ and $97.98\pm1.06\%$, respectively. The optimal hitting occurs at a polynomial time both for CTQWs and CRWs. However, the optimal hitting efficiency of CRWs falls exponentially with layer depth, in contrast with the polynomial decrease tendency of CTQWs, verifying an exponential quantum speedup over the hitting performance by CRWs.
We also demonstrated quadratically fast hitting performance on nets and grids with up to 210 vertices, which are the largest-scale experimental demonstrations of fast hitting dynamics up to now \cite{SP}.

Another essential feature of QW dynamics is known as mixing \cite{Aharonov2001}. Although unitary QWs hardly converge to stationary distributions as CRWs, one can capture a dynamically stabilized situation by observing the long-time average distribution of quantum walkers on the graphs.
Of much significance to the heart of quantum speedups for many QW-based algorithms \cite{Wocjan2008} is quantum mixing time \cite{Chakraborty2020}, that is, the minimum time required for a QW to converge close to its average limiting distribution.
However, the demands for long-time stability and intensively repeated measurements remain challenges for the experimental investigation. 
From the overall mixing process depicted by our system, we observed nearly quadratic speedups of CTQW mixing over CRW on eNets (Fig.\ref{fig:fig2}C) and eGrids (Fig.\ref{fig:fig2}D). Compared to CRWs, CTQWs on the 20-layer net and the 36-layer grid almost halve in the evolution time for mixing. For the mixing on the largest-scale net, the average of the fidelities between the experimental and theoretical probability distributions is $96.28\pm1.86\%$, and the similarity between measured and theoretical limiting distribution reaches $99.76\%$. See more details in the supplementary materials \cite{SP}.
As far as we know, these are the first experimental demonstrations of QW mixing dynamics.

Additionally, we implemented four QW-based applications on large-scale graphs using our system.
(1) \emph{Centrality measure}. 
Based on QWs, quantum-enhanced algorithms were proposed for ranking the vertex centrality of graphs and further used for large-scale network analysis \cite{Izaac2017a}.
We performed a CTQW-based centrality measure algorithm on eScale-free random networks. Fig.\ref{fig:fig3}A presents the experimentally obtained CTQW-based centrality results of a 55-vertex eScale-free network, which correlates well with its eigenvector centrality (similarity = 95.68\%). This is the largest-scale experimental realization of CTQW-based centrality measure to date that validates the potential of the QW in large-scale network analysis.
(2) \emph{Search on networks}. Finding marked vertices in a graph can be solved in the framework of QWs. It has been proved that CTQWs can search on almost all graphs of size $N$ in time $\mathcal{O}(\sqrt{N})$ and thus provide quadratic speedup over classical algorithms \cite{Chakraborty2016}.
We benchmarked the performance of a CTQW-based search algorithm \cite{Qiang2021} on 1,000 randomly eErd\"{o}s-R\'{e}nyi networks with sizes ranging from 15 to 210. With the experimental data statistics, we show the optimal search time starting from three vertices to find the other three marked vertices scales as $0.8026\sqrt N+4.06895$ (Fig.\ref{fig:fig3}B) and experimentally demonstrate the quadratic speedup for the first time.
(3) \emph{Graph isomorphism test}. Another application of QWs is to tackle the graph isomorphism problem, that is, determining whether two given graphs are isomorphic (two graphs are isomorphic if one can be obtained from the other by relabeling the vertices). To demonstrate a CTQW-based algorithm \cite{Qiang2021} on graphs with 210 vertices, we constructed the graph certificates from experimentally obtained CTQW evolution results. By comparing the total variance distance of graph certificates between graphs, their isomorphism is distinguished. As shown in Fig.3C, the experimental and theoretical results are highly consistent. For the isomorphic graph pair, the average distance of graph certificates is close to 0, while non-isomorphic graphs achieve a much larger distance. 
(4) \emph{Topological phase simulation}. By encoding the property of particles and geometry of graphs, QWs can be used to model a wide variety of physical systems and processes. As shown in Fig.\ref{fig:fig3}, we demonstrated the topological invariants of two typical higher-order
topological insulators, the 2D Su-Schrieffer-Hegger (SSH) model \cite{Liu2017} and Benalcazar-Bernevig-Hughes (BBH) model \cite{Benalcazar2017}, by probing the long-time averaged values of extended mean chiral displacement (AMCD) \cite{Maffei2018} and mean chiral quadrupole moment (AMCQM) \cite{Mizoguchi2021} in single-particle and two-fermion CTQWs, respectively. 
The bulk topology of the 2D SSH model should be characterized by two topological invariants along the $x$ and $y$ dimensions. The experimental asymptotic values of AMCD in topological phases of the 2D SSH model are $(0.470\pm0.002, 0.498\pm0.001)$, which are obviously distinct from the values in trivial  phases, $(-0.046\pm0.004, 0.013\pm0.003)$. In Fig.\ref{fig:fig3}, we present the results along the $y$ dimension as an example. Similarly, the experimental asymptotic values of AMCQM of the BBH model are $0.513\pm0.002$  in topological phases and $0.014\pm0.003$ in trivial 
phases.   
All these demonstrations of applications showcased the potential of our system for practical large-scale network analysis and the field of many-body system quantum simulations.

In conclusion, we have designed and realized a full-stack photonic quantum computing system for simulating universal large-scale QWs and their applications.
It allows investigating unique QW dynamics features on large-scale graphs, where we experimentally demonstrated the exponential quantum speedup in hitting and the quadratically quantum speedup in mixing for the first time.
Using the system, we also demonstrated versatile applications in high precision, from graph-theoretic applications and quantum simulations of topological
phases. 
Our work shows that the dedicated integrated photonic system with particular QW models paves a viable path to bring quantum photonics to fruition in practical applications. 
With the rapid development of integrated quantum photonics \cite{Politi2008,Qiang2018,Bartolucci2021a}, such quantum photonics-enabled computers would be accelerated to achieve practical quantum advantages.

\bibliography{apstemplate}

\ifarXiv
    

\section*{Supplementary material}

# Supplementary Materials

# Contents

# 1. Models and methods of quantum walk simulation

## 1.1 Single-particle CTQW

In this report, we mainly focus on the continuous model of quantum walk, namely, the continuous-time quantum walk (CTQW). The formalism of CTQW [6] is introduced closely from its classical counterpart, the continuous-time random walk (CTRW). CTRW is a Markov process where a walker walks on a graph $G$ with no time restrictions, i.e., the walker evolves with an arbitrary long walking time. Suppose the graph $G$ has a set of vertices $V = \{v_1, v_2, \ldots, v_N\}$, one vertex $v_a$ can be connected to the other vertex $v_b$ by an edge $(v_a, v_b) \in E$. The adjacent matrix $A$ of the graph $G$ is defined by

$$A_{a,b} = \begin{cases} 1, & (v_a, v_b) \in E \\ 0, & (v_a, v_b) \notin E. \end{cases} \tag{1}$$

The probability of walker jumping from one vertex to any adjacent vertex in a time $\varepsilon$ is $\gamma\varepsilon$, where $\varepsilon \to 0$, $\gamma$ is the jumping rate and we generally set $\gamma = 1$. The stochastic generator matrix $M$ is defined by

$$M_{a,b} = \begin{cases} -\gamma A_{a,b}, & a \neq b \\ \gamma D(a), & a = b, \end{cases} \tag{2}$$

where $D(a)$ denotes the degree of the vertex $v_a$. Let $\mathbf{P}(t)$ represents the probability distribution of the walker at time $t$. Then the master equation governing the time evolution of CTRW is given by

$$\frac{d}{dt}\mathbf{P}(t) = -M\mathbf{P}(t). \tag{3}$$

Thus, the evolution of CTRW can be described by

$$\mathbf{P}(t) = e^{-Mt}\mathbf{P}(0). \tag{4}$$

For CTQW on a graph $G$, let $|\psi(t)\rangle$ denotes the quantum state of the walker, and the Hilbert space is spanned by the localized quantum states at the vertices $\mathcal{H} = \text{span}\{|1\rangle, |2\rangle, \ldots, |n\rangle\}$. The dynamics of CTQW are governed instead by the Schrödinger equation,

$$i\hbar \frac{d}{dt}|\psi(t)\rangle = H|\psi(t)\rangle, \tag{5}$$

where $H$ is the Hamiltonian, and we use $H = \gamma A$ in our report. For simplicity, we set $\hbar = 1$ and $\gamma = 1$. Then the general solution to the CTQW system is

$$|\psi(t)\rangle = U(t)|\phi_0\rangle, \tag{6}$$

where

$$U(t) = e^{-iHt} \tag{7}$$

is the unitary evolution operator with specific evolution time $t$, and $|\phi_0\rangle = |\psi(0)\rangle$ is the initial state. The probability that the walker is localized at the vertex $v_w$ is given by

$$P_w^{|\phi_0\rangle}(t) = |\langle w|U(t)|\phi_0\rangle|^2. \tag{8}$$

## 1.2 Simulating multiple-particle CTQW with control over particle properties using entangled photons

Here we review an entanglement-driven scheme to simulate the CTQW evolutions of multiple particles with tunable exchange statistics and indistinguishability. More details can be found in [39] and [20].

First consider the CTQW of two indistinguishable non-interacting particles with exchange statistics denoted as $\phi$ ($\phi = 0$ for bosons and $\phi = \pi$ for fermions). When two particles start from vertices $j$ and $k$ of the graph $G$, the initial state is given by $a_j^\dagger a_k^\dagger |0\rangle$. After evolution time $t$, the correlated detection probability of the two particles being at $r$ and $q$ is

$$\Gamma_{r,q}^\phi = |\langle 0|a_r a_q U a_j^\dagger a_k^\dagger |0\rangle| = |U_{r,j}U_{q,k} + e^{i\phi}U_{r,k}U_{q,j}|^2, \tag{9}$$

where $U = U(t)$, $a_j^\dagger$ and $a_j$ are the creation and annihilation operators at the vertex $j$, respectively.

Now consider the case of two particles with extra partial indistinguishability $\gamma$ ($0 \leq \gamma \leq 1$), which yields the initial state as $a_{j,h}^\dagger \left(\gamma a_{k,h}^\dagger + \sqrt{1-\gamma^2} a_{k,v}^\dagger\right)|0\rangle$, where $h$ and $v$ represent the horizontal and vertical polarizations, respectively. Projecting on the state

$1/\sqrt{2}\langle 0|a_{r,h}(a_{q,h}+a_{q,v})$, the correlated detection outcome of the two particles at $r$ and $q$ is given by

$$\Gamma_{r,q}^{\phi,\gamma} = \left|\frac{1}{\sqrt{2}}\langle 0|a_{r,h}(a_{q,h}+a_{q,v})Ua_{j,h}^{\dagger}\left(\gamma a_{k,h}^{\dagger}+\sqrt{1-\gamma^2}a_{k,h}^{\dagger}\right)|0\rangle\right|^2$$
$$= \frac{1}{2}\left|\gamma\left(U_{r,j}U_{q,k}+e^{i\phi}U_{r,k}U_{q,j}\right)+\sqrt{1-\gamma^2}U_{r,j}U_{q,k}\right|^2. \tag{10}$$

and $\Gamma_{r,q}^{\phi,\gamma}$ with arbitrary $r$ and $q$ presents the complete probability distribution. Varying $\gamma$ and $\phi$ allows continuous control over the particles from indistinguishability ($\gamma=1$) to distinguishability ($\gamma=0$) and from Bose-Einstein ($\phi=0$) to Fermi-Dirac ($\phi=\pi$) quantum interference, respectively.

Consider a two-photon state

$$|\psi(\phi,\theta)\rangle = \left(e^{i\phi}\sin(\theta)a_1^{\dagger}b_2^{\dagger}+\cos(\theta)a_2^{\dagger}b_1^{\dagger}\right)|0\rangle, \tag{11}$$

where $\theta$ and $\phi$ are the relative proportion and phase of entanglement components. With $\theta$ varying from 0 to $\pi/4$, $|\psi(\phi,\theta)\rangle$ covers the full spectrum from a fully separate state to a maximally entangled state. By sending each of the entangled photons through two identical copies ($U^a$ and $U^b$) of an evolution process $U$, the state evolves to

$$|\psi(\phi,\theta,U^a,U^b)\rangle = \sum\left(e^{i\phi}\sin(\theta)U_{s,1}^a U_{t,2}^b+\cos(\theta)U_{s,2}^a U_{t,1}^b\right)a_s^{\dagger}b_t^{\dagger}|0\rangle. \tag{12}$$

Then the correlated detection probability at the mode $r$ of $U^a$ and $q$ of $U^b$ can be obtained as

$$P_{r,q}^{\phi,\theta} = \left|e^{i\phi}\sin(\theta)U_{r,1}^a U_{q,2}^b+\cos(\theta)U_{r,2}^a U_{q,1}^b\right|^2. \tag{13}$$

Compare (10) and (13), by controlling the generated entangled state (with $\theta$ and $\phi$) and the unitary transformations ($U^a$ and $U^b$), we can simulate the correlated detection $\Gamma_{r,q}^{\phi,\gamma}$ using two-photon interference $P_{r,q}^{\phi,\theta}$

$$\Gamma_{r,q}^{\phi,\gamma} = \mu P_{r,q}^{\phi,\theta}, \tag{14}$$

where $\mu = \frac{1}{2}\left(1+\gamma^2 + 2\gamma\sqrt{1-\gamma^2}\right)$ and $\theta = \arctan\frac{\gamma+\sqrt{1-\gamma^2}}{\gamma}$.

Therefore, by sending each particle of an entangled two-photon pair, passing through identical copies of a CTQW evolution, and then measuring the correlated detection probabilities, we can simulate two-particle CTQW with tunable particle indistinguishability, exchange symmetry, and underlying CTQW evolution. We can further take complete control of particle indistinguishability, exchange symmetry, and underlying CTQW evolution by governing the degree of the entanglement, relative phase, and unitary transformation, respectively. Extended data Fig. S7 gives a result of the experimental simulation of CTQW of two correlated particles on a 20-vertex circle with tunable indistinguishability and exchange symmetry at the evolution time π/8.

This entanglement-based approach also allows scalable simulation of multiple indistinguishable particle interference, by harnessing multi-partite entanglement [39]. Consider $P$ identical particles with exchange statistics $\phi$, which are initially located at vertices of the graph $G: \vec{v} = \{v_1, v_2, \ldots, v_P\}$. After the evolution of CTQW, the correlated detection outcome of the two particles at the output $\vec{o} = \{o_1, o_2, \ldots, o_P\}$ is given by

$$\Gamma_{\vec{o}}^{\phi} = \left| \sum_{\sigma_v \in S_P} e^{i\tau(\sigma_v)\phi} \prod_{j=1}^{P} A_{o_j, \sigma_{v_j}} \right|^2, \qquad (15)$$

where $\sigma_v$ represents an element of the permutation group $S_P$ acting on $\vec{v}$, $\tau(\sigma_v)$ represents the minimum number of neighboring swaps that maps $\sigma_v$ to $\vec{v}$ and $\sigma_{v_j}$ is $j$-th element of $\sigma_v$. By subjecting the generalized $P$-partite, $P$-level entangled state

$$\frac{1}{\sqrt{P!}} \sum_{\sigma_v \in S_P} e^{i\tau(\sigma_v)\phi} \prod_{j=1}^{P} a_{\sigma_{v_j}}^{(j)\dagger} |0\rangle \qquad (16)$$

to $P$ copies of the evolution process $U$ with one particle $a^{(j)\dagger}$ injected into each copy $U^{(j)}$, the quantum interference in (15) can be obtained through the correlated detection probability of $P$ particles output from $o_1$ of $U^{(1)}$, $o_2$ of $U^{(2)}$, $\cdots$, $o_P$ of $U^{(P)}$ (with a $1/P!$ pre-factor).

## 1.3 Simulating single-particle CTQW on exponentially sized graphs from multi-particle interference

In this section, we show how to simulate single-particle CTQW on exponentially sized graphs with polynomial sources, by harnessing the multi-particle quantum interference on small graphs. More details and further discussion can be found in [20-22].

We first extend the CTQW to $P$ non-interacting and fully distinguishable particles on the graph $G$ with $N$ vertices, then decompose the full Hamiltonian in the $N^P$-dimensional Hilbert space as

$$H_D^{(P)} = H^{\oplus P}, \qquad (17)$$

where the first term is the Kronecker sum of $P$ free-particle Hamiltonian on the graph $G$, namely $H = A$. Thus, the single-particle CTQW on a $N^P$-vertex Cartesian product graph $G_D^{(P)}$, of which the adjacency matrix is $A_D^{(P)} = A^{\oplus P}$, can be simulated by $P$ fully distinguishable particles evolution on the graph [20].

However, when the particles involved in the CTQW are fully indistinguishable, we can no longer distinguish between states that particles occupy the same sets of vertices. For CTQW of $P$ fermions, the anti-symmetrized basis state $|j_1, j_2, \ldots, j_P\rangle_F$ ($j_1 < j_2 < \ldots < j_P$) represents $P$ fermions localized at vertices $j_1$ through $j_P$ and excludes states that multiple particles occupy the same vertices. Whereas, for CTQW of $P$ bosons, the symmetrized basis state $|j_1, j_2, \ldots, j_P\rangle_B$ ($j_1 \leq j_2 \leq \ldots \leq j_P$) represents $P$ bosons localized at vertices $j_1$ through $j_P$, where these vertices need not to be distinct.

By expressing the $P$-particle Hamiltonian $H^{\oplus P}$ in the allowed particles-on-vertices basis [22], we obtain a reduced-dimension Hamiltonian $H_{reduced}^{(P,\phi)}$ without any redundant states,

$$\left(H_F^{(P)}\right)_{i,j} = {}_F\langle i_1, i_2, \ldots, i_P | H^{\oplus P} | j_1, j_2, \ldots, j_P\rangle_F, \qquad (18)$$

$$\left(H_B^{(P)}\right)_{i,j} = {}_B\langle i_1, i_2, \ldots, i_P | H^{\oplus P} | j_1, j_2, \ldots, j_P\rangle_B, \qquad (19)$$

where $i$ and $j$ denote the order of basis states in all allowed bases, respectively. In fact, these reduced Hamiltonians form the adjacency matrix of the state transition graphs $G_F^{(P)}$ and $G_B^{(P)}$, of which the vertices represent the basis state space of $P$-fermion or $P$-boson CTQW on the graph $G$, and the edges represent the allowed transfer paths. The dimension of reduced fermionic/bosonic graph is $\binom{N}{P}$ and $\binom{N+P-1}{P}$, respectively. Rather than directly simulating single-particle CTQW on some exponentially sized graphs, we can instead simulate the dynamics via a multi-particle CTQW on a much smaller scale graph.

Ref. [21] presented a method of constructing the reduced-dimension adjacency matrix for a multiple fermionic CTQW. Here we present an example to illustrate the construction of a reduced bosonic graph of multiple bosons noninteracting CTQW. The appropriate basis states for 2-boson CTQW can be

$$|j_1, j_2\rangle_B = \begin{cases} \frac{1}{\sqrt{2!}} \sum_{\sigma \in S_2} |j_{\sigma(1)}, j_{\sigma(2)}\rangle, & j_1 \neq j_2 \\ |j_1, j_2\rangle, & j_1 = j_2, \end{cases} \quad (20)$$

where $\sigma$ is a member of the permutation group $S_2$, representing all permutations acting on $\{j_1, j_2\}$. Then the bosonic reduced-dimension Hamiltonian $H_B^{(2)}$ is given by

$$\left(H_B^2\right)_{i,j} = {}_B\langle i_1, i_2 | H \oplus H | j_1, j_2 \rangle_B. \quad (21)$$

$H_B^{(2)}$ is also the adjacency matrix of the state transition graph $G_B^{(2)}$. We call the constructed graph as the bosonic graph for two particles holding bosonic exchange symmetry.

As shown in extended data from Fig. S8 to Fig. S11, we implemented two-boson CTQW on randomly glued trees (RGTs), hypercubes (Cubes), circles, lines, Scale-free networks, and Erdős-Rényi networks to construct larger graphs with up to 36 layers and 210 vertices. The constructed graphs are named as extended RGTs (eRGTs in short), extended Cubes (eCubes), extended nets (eNets), extended grids (eGrids), extended Scale-free (eScale-free) networks, extended Erdős-Rényi (eErdős-Rényi) networks, respectively. In our report, the layer depth of the graph denotes the maximin one-way moving step from one vertex to any other vertex. We

experimentally implemented single-particle CTQWs on these constructed graphs, to explore the fast hitting and mixing dynamics and demonstrate various CTQW-based algorithms.

### 1.4 Analysis of complexity and scalability

This entanglement-driven scheme has polynomial cost in physical resources for even the classically intractable task such as $P-$boson quantum walks in $N$ mode unitary process by harnessing multi-partite entanglement [20]. A scaled version of our system can simulate universal quantum walks of $P$ correlated particles by injecting a $P-$partite entangled photonic state into $P$ copies of $N-$dimensional programmable linear optical circuits. This requires physical resources as $P$ photons and $P$ copies of $N-$dimensional reconfigurable optical circuits, together with a negligible number of phase shifters to control the $P-$partite entangled photonic state. Note that each $N-$dimensional reconfigurable optical circuit consists of $N(N-1)/2$ phase shifters and some fixed components.

The $N-$dimensional underpinning unitary evolution operators are classically computed on the master microcomputer, which approximately costs $\mathrm{O}(N^3)$. Then the configuration of all $P$ copies of $N-$dimensional reconfigurable optical circuit costs $PN(N-1)/2$. Therefore, $P$ correlated particles quantum interference can be simulated on a scaled system with only an overall polynomial complexity of $\mathrm{O}(N^3+PN^2)$. In addition, our device allows to investigate the dynamics and verify the performance of quantum walks on exponentially-sized graphs (via multi-boson walk) with polynomial cost.

## 2. System details

As shown in Fig.1, the system is composed of a software stack and a hardware stack. The hardware stack consists of

(a) a packaged photonic chip module, the suitable and stable microenvironment where a programmable silicon photonic chip works,

(b) a pump module producing amplified laser for photon-pair generation,

(c) an electrical control module that follows driver instructions and sends electric signals to configure the chip, and

(d) a single-board microcomputer running the layered software stack to compile applications into hardware operations and the device driver to master all submodules.

These modules are compactly contained within an 85cm×60cm×55cm portable case. Note that the detection system is peripheral because it is computational power independent, and users have the freedom to configure the amount and organization of detectors.

### 2.1 Programmable silicon photonic chip

The core of the system is a programmable silicon photonics chip capable of generating the entangled state as (12) and applying unitary evolutions. The chip monolithically embeds more than 900 components, including two spiral-waveguide spontaneous four-wave mixing (SFWM) photon-pair sources, 228 thermo-optic phase shifters, 150 multimode interferometers (MMIs), 3 waveguide crossers, 64 optical grating couplers, and 464 electrical pads. We place one phase shifter at both Mach-Zehnder interferometer (MZI) arms to guarantee the fabrication uniformity, thus improving the accuracy of linear optical circuit operations. Half of these phase shifters work simultaneously, and others stay as spares in case of electrical damage.

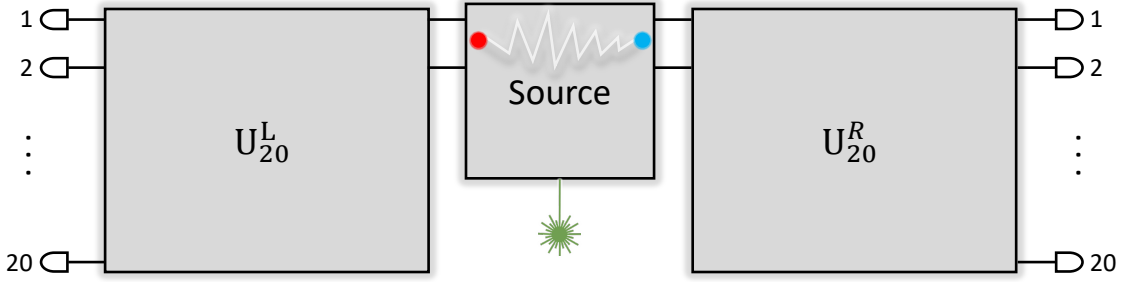

**Fig. S1 Schematic of the programmable silicon photonic chip.** The chip mainly consists of on-chip entangled photon sources and two universal 20-dimensional linear optical circuits. The generated two-photon states and implemented 20-dimensional unitary transformations can all be programmed by tunning on-chip thermo-optic phase shifters.

The pump laser is launched into the middle of the chip through on-chip grating couplers, and split across the two SFWM sources $S_L$ and $S_R$ with adjustable ratio and relative phase continuously tuned by an MZI and an extra phase shifter, respectively. The two SFWM sources are coherently pumped to create possible signal-idler photon pairs. Then two nondegenerate photons are stochastically routed to either of two universal linear optical circuits $\text{LOC}_L$ and $\text{LOC}_R$. Post-selecting the cases that the idler photons exit at the left circuit and the signal photons exit at the right, yields the path-entangled two-photon state as

$$|\psi(\phi,\theta)\rangle = \left(e^{i\phi}\sin(\theta)a_1^\dagger b_2^\dagger + \cos(\theta)a_2^\dagger b_1^\dagger\right)|0\rangle, \qquad (22)$$

with a success probability of $1/4$, where $a_i^\dagger$ and $b_j^\dagger$ are creation operators associated to signal photon at $i$ mode and idler photon at $j$ mode, respectively. Each 20-mode optical circuit consists of MMIs and phase shifters arranged in the triangular configuration [40]. They can implement arbitrary 20-dimensional unitary transformations with fixed input mode, denoted as $U^L$ and $U^R$. Through the optical circuits, the output quantum state of the two entangled photons in the 400-dimensional Hilbert space evolves into

$$|\psi(\phi,\theta,U^L,U^R)\rangle = \sum_{1\leq s,t\leq 20}\left(e^{i\phi}\sin(\theta)U_{s,1}^L U_{t,2}^R + \cos(\theta)U_{s,2}^L U_{t,1}^R\right)a_s^\dagger b_t^\dagger|0\rangle. \qquad (23)$$

Then the correlated detection probability at the mode $r$ of $U^L$ and $q$ of $U^R$ can be obtained as

17

$$\mathbf{P}_{r,q}^{\phi,\theta} = \left| e^{i\phi} \sin(\theta) U_{r,1}^{L} U_{q,2}^{R} + \cos(\theta) U_{r,2}^{L} U_{q,1}^{R} \right|^2. \tag{24}$$

**2.2 Packaged photonic chip module**

The chip is optically aligned and then firmly glued with a 64-channel V-groove fiber array (VGA) using the optical adhesive so that the system shuffles off energy-draining manual alignment with precise nano-positioning stages or maintenance relying on a vibration isolation workstation. By measuring the insertion loss on several on-chip alignment waveguide loops, the average coupling loss is estimated as ~5.0 dB/facet.

All 464 electrical pads are placed on three sides of the chip in a staggered arrangement. Each pad is wire-bonded to a multi-layered printed circuit board (PCB), with vulnerable gold wires protected by solid insulation gum. Cables connect the PCB with a self-developed 512-channel, 16-bit precision electrical control module (ECM). ECM can respond to the device driver and send the electrical signals to reconfigure all the on-chip thermo-optic phase shifters parallelly in $\leq 0.5$ milliseconds.

When all phase shifters work simultaneously, the thermal power of the chip fluctuates at a kHz rate and reaches up to 35 Watts, as high as the Thermal Design Power of Intel Core i9-12900T CPU. The phase shifters rely on the thermal-optic effect; thus, the chip requires to be actively held at a constant temperature, regardless of internal heat generation and external temperature fluctuations. An active temperature stabilizer module capable of rapid heating and cooling is essential. We first tightly attach the chip to a copper sub-mount, which acts as a heat reservoir with effective thermal conductivity. A thermistor embeds into the sub-mount just below the chip to probe the actual and some recent temperatures. Then we clamp this combination to a thermoelectric cooler (TEC), with a cooling capacity of 120 Watts. The TEC can pump heat from its upper surface to a coolant circulatory module below. We fill the apertures between all parts with high heat-conductive thermal adhesive. A TEC controller will compare the current chip temperature, then produces electric current in order to achieve the desired temperature using Proportion Integration Differentiation (PID) optimization. With this closed-loop control module,

the chip remains overall stable at the setting temperature (18.0℃) with fluctuation of less than ±0.03 ℃, even when all phase shifters simultaneously work.

We encapsulate this optically, electrically, and thermally packaged chip into a 38cm×35cm ×17cm compact board, with 64 optical ports and 464 electrical ports left for photon and electrical I/O. Complete packaging moves photonic chips out of the demanding laboratory and works leisurely in the more mundane environment. Although suffering thousands of kilometers of motor transport, this module still works well.

## 2.3 Laser & Amplifier

This module comprises a continuous-wave laser tuned to a wavelength of 1549.32 nm, an erbium doped optical fiber amplifier (EDFA), a dense wavelength division multiplexing (DWDM) filters, and a polarization controller (PC). After this module, approximately 200 mW of bright light is collected and then injected into the chip for photon-pair generation.

## 2.4 Electrical control module

Our electrical control module (ECM) is designed with the master-slave mode to support the parallel configuration of electrical signals. The digital signal control board (DSCB) manufactured based on FPGA can associate with four analog voltage control sub-boards (AVCSBs). Each AVCSB has 128 voltage output channels with 16-bit precision. The delay of communication and configuration is limited to 0.5 milliseconds, which means the ECM can simultaneously apply voltage with 512 channels in a short time.

## 2.5 Single-board computer

The computing capability of this system is mainly provided by the programmable silicon photonics chip; thus, a microcomputer capable of compiling CTQW algorithms and coordinating control and measurement modules, rather than a server computer, is competent enough. We employ a Raspberry Pi 4 Model B, a three-inch single-board computer, to drive software stacks

and all the hardware components. The systems are also available from any remote computer with an internet connection to Raspberry Pi.

**2.6 Software stack**

To support executing various tasks in a user-friendly way, we developed a layered software stack as the bridge between the abstract quantum algorithms and operations in hardware. The software stack is divided into three abstraction layers: Quantum algorithms, Quantum walk interface, and Device drivers. At the highest level, quantum algorithms, coded in Julia programming language, are expected to be modeled with the execution of QWs. At the middle level, the quantum walk interface translates abstraction algorithms to the corresponding settings of QWs, such as particle property, initial state, evolution time, and underlying graph structure. Then the interface compiles QW settings into chip configurations and hardware operations. The device drivers interact with the chip via optical and electrical ports, and perform three prominent functions, that is, pumping by an amplified laser, continuous and parallel configuration of all the on-chip phase shifters by a self-developed 512-channel, 16-bit precision electrical control module, and detection of the chip by photon detectors.

# 3. Basic performance test

## 3.1 Performance of linear optical circuits

We verify the reconfigurability and precision of the 20-mode optical circuits by generating and implementing 10,000 Haar-random 20-dimensional unitary operations on the device. We measure the output optical intensity distribution for each input mode of each circuit. For each case, the classical fidelity $F_c$ between the experimentally obtained distribution and the theoretical result is

$$F_c = \sum_{i=1}^{20} \sqrt{P_{\exp,i} P_{th,i}}. \tag{25}$$

where $P_{\exp}$ and $P_{th}$ represent the experimental and theoretical output optical distribution values. The statistical fidelities are as high as $99.49 \pm 0.14\%$, $99.51 \pm 0.14\%$, $99.60 \pm 0.10\%$, and $99.20 \pm 0.30\%$ for $\text{LOC}_L$ with input mode 1, $\text{LOC}_L$ with input mode 2, $\text{LOC}_R$ with input mode 1, and $\text{LOC}_R$ with input mode 2, respectively. The histograms of measured fidelities for each case are shown in Fig. S2.

## 3.2 Performance of high-dimensional quantum states generation

We performed experiments on the bipartite $d$-dimensional entangled system, generated by employing two SFWM sources and $d$ of the 20 modes of each linear optical circuits. After randomly generating the initial entangled state and implementing $d$-dimensional Haar-random unitary operations, the two-photon coincidence measurements were collected between the $d$ modes of two circuits. We define the classical statistic fidelity $F_S(d^2)$ of the $d^2$-dimensional quantum state as

$$F_S(d^2) = \sum_{1 \leq i,j \leq d} \sqrt{P_{i,j}^{\exp} P_{i,j}^{th}}, \tag{26}$$

where $P^{\exp}$ and $P^{th}$ represent the experimental and theoretical correlated detection probability given by (24). Histograms of classical statistic fidelity $F_S(d^2)$ for $d^2$-dimensional quantum state ($d = 2,4,6,\ldots,20$) are presented in Fig. S3.

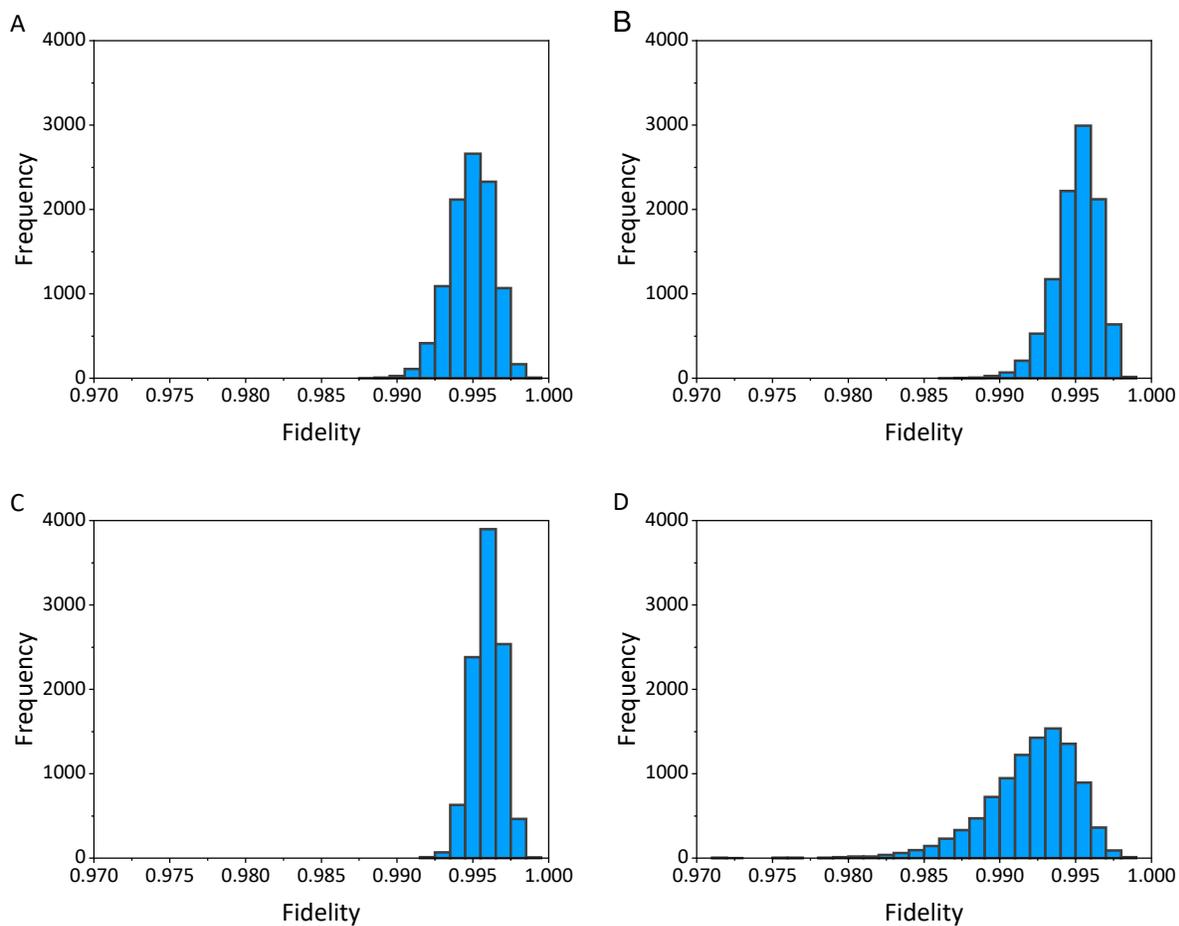

**Fig. S2 Histograms of the measured classical fidelities for 10,000 Haar-random unitary operations. (A)** Random unitary operations implemented in $\mathrm{LOC}_L$ with input mode 1. **(B)** Random unitary operations implemented in $\mathrm{LOC}_L$ with input mode 2. **(C)** Random unitary operations implemented in $\mathrm{LOC}_R$ with input mode 1. **(D)** Random unitary operations implemented in $\mathrm{LOC}_R$ with input mode 2.

22

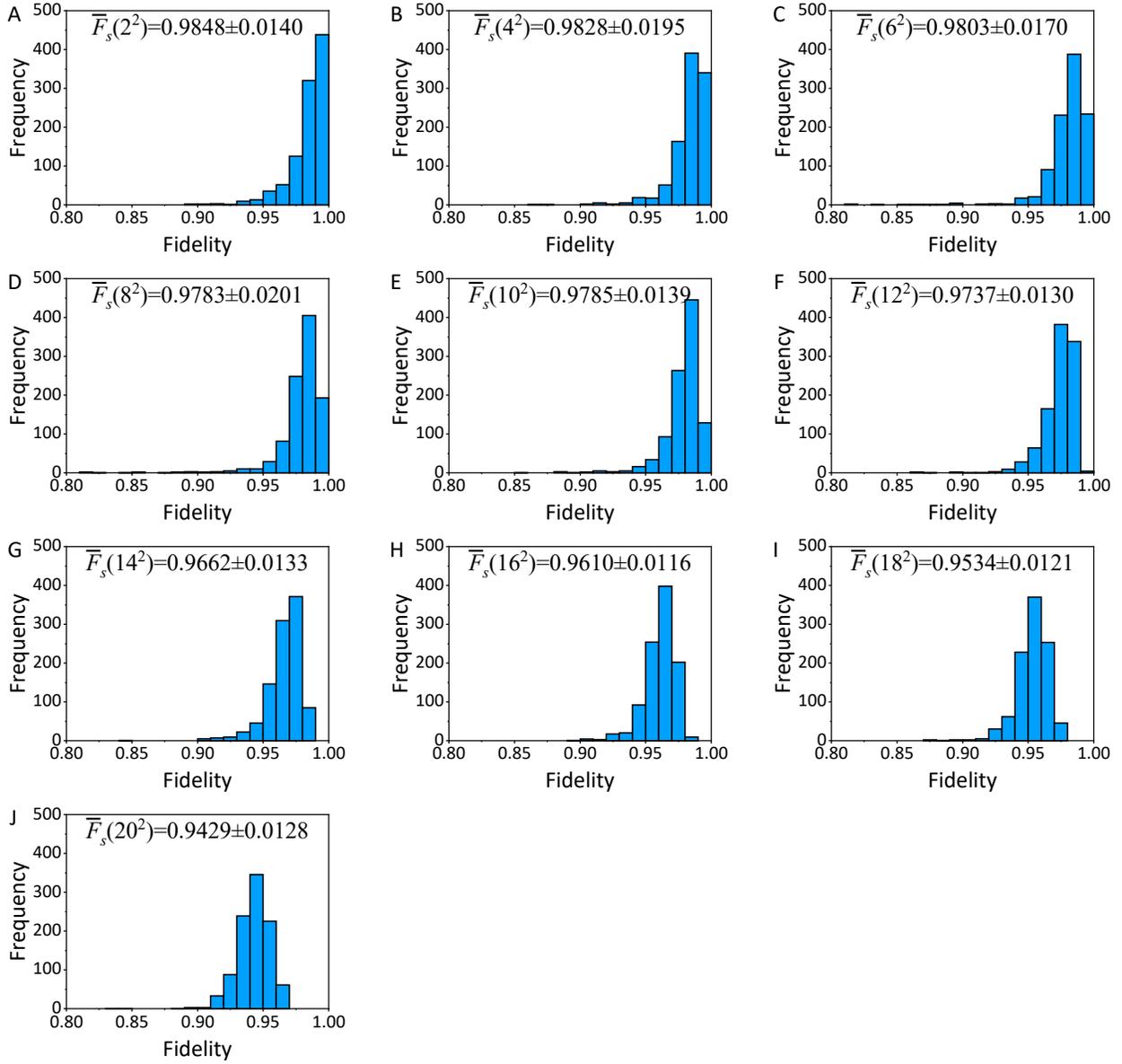

**Fig. S3 Histograms of the measured classical statistic fidelities for quantum states in the Hilbert space with dimensions from 4 to 400 (A)-(J).**

### 3.3 Heavy output generation test

We generalize the heavy output generation problem, which is widely used in the quantum volume metric for benchmarking quantum computers [41], to validate the largest size of quantum states that QW2020 successfully implements. Given the proportion $\theta$ and relative phase $\phi$ of entanglement components (hereinafter associated with particle properties), and model circuits $U^L$ and $U^R$ (hereinafter associated with CTQW evolutions), the ideal output distribution in (24) can be rewritten as

$$p_{\phi,\theta,U^L,U^R}(x) = \left| \langle x | \sum_{1 \leq s,t \leq 20} \left( e^{i\phi} \sin(\theta) U^L_{s,1} U^R_{t,2} + \cos(\theta) U^L_{s,2} U^R_{t,1} \right) a_s^\dagger b_t^\dagger |0\rangle \right|^2, \quad (27)$$

where $x \in \{1, 2, \cdots, 400\}$ is a coincidence detection sequence number. For instance, for coincidence detection between $a$-th port of linear optical circuits $\text{LOC}_L$ and $b$-th port of $\text{LOC}_R$, then $x = 20a + b$. The set of output probabilities given by $p_{\phi,\theta,U^L,U^L}(x)$ are sorted in ascending order $p_1 \leq p_2 \leq \cdots \leq p_{400}$, and the median of the set of probabilities is $p_{med}$. The heavy outputs are

$$H_{\phi,\theta,U^L,U^R} = \left\{ x \in \{1, 2, \cdots, 400\} \text{ such that } p_{\phi,\theta,U^L,U^R}(x) > p_{med} \right\}. \quad (28)$$

Then the heavy output generation problem is generalized to choose a set of coincidence detection sequence numbers such that more than two-thirds are heavy. For an ideal device, the expected heavy output probability is asymptotically ~0.85, while it drops down to ~0.5 if the device is completely depolarized.

After implementing the configuration of the chip, we can experimentally observe a probability distribution as $q_{\phi,\theta,U^L,U^R}(x)$, then the probability of sampling a heavy output is

$$h_{\phi,\theta,U^L,U^R} = \sum_{x \in H_{\phi,\theta,U^L,U^L}} q_{\phi,\theta,U^L,U^R}(x). \quad (29)$$

If the estimated $h_{\phi,\theta,U^L,U^R} > \frac{2}{3}$ is confidently guaranteed, we can say that the achievable largest dimension of the Hilbert space reaches 400. The pseudo algorithm for $h_{\phi,\theta,U^L,U^R}$ estimation is

presented in Algorithm 1. Experiments were repeated with $n_c = 1000$ model circuits, and the sample number $n_s$ was set as $> 10^6$ to ensure a strict confidence, interval as further described in [41]. Experimentally estimated heavy output probabilities for quantum states generated by QW2020 is 0.8101. Thus, we conclude the largest dimension of available Hilbert space reaches 400, with a 97.5% one-sided confidence interval.

---

**Algorithm 1** Check heavy output generation

---

**Input:** number of experiments $n_c$, threshold coincidence count $n_s$

1: $n_h \leftarrow 0$
2: **for** $i = 1 : n_c$ **do**
3:     $\theta, \phi \leftarrow$ Haar random entanglement parameter
4:     $U^L, U^R \leftarrow$ Haar random 20-diemensional unitary
5:     calculate $H_{\phi,\theta,U^L,U^R}$ by classical simulation
6:     configure chip to implement $\theta, \phi, U^L, U^R$
7:     **while** coincidence count $< n_s$ **do**
8:         perform coincidence detection
9:     **end**
10:     calculate $q_{\phi,\theta,U^L,U^R}(x)$ from detected coincidence records
12:     calculate $h_{\phi,\theta,U^L,U^R}$ from $q_{\phi,\theta,U^L,U^R}(x)$
13:     **if** $h_{\phi,\theta,U^L,U^R} > 2/3$ **then do**
14:         $n_h \leftarrow n_h + 1$
15:     **end**
16: **end**
17: **return** $\dfrac{n_h - 2\sqrt{n_h(n_s - n_h/n_c)}}{n_c n_s} > \dfrac{2}{3}$

## 4. Details for quantum walk hitting

The evolution pattern and hitting efficiency of CTRW are obtained by the formalism (4). The jump rate in CTRW simulations, CTQW simulations, and CTQW experiments are all set as 1. For CTQWs, which intrinsically yield non-stationary distribution, the optimal hitting occurs when the measured probability of the walker at the exit reaches its first peak value. While for classical hitting, the probability approaches asymptotically to a stationary distribution. We consider that classical optimal hitting occurs when the hitting efficiency has a deviance of no more than 1% from optimal hitting efficiency.

More experimental details in the simulations of exponentially fast hitting dynamics of CTQW on eRGTs, eCubes, eNets, and eGrids are listed in Table S1. For eRGTs with layer depth $N$ up to 10 and eCubes with $N$ up to 8, comparisons between the CTQWs and CTRWs in the optimal hitting efficiency and hitting time are presented in experimental data and theoretical prediction in Table S2, Fig. S12, and Fig. S13. The experimental results of optimal hitting efficiency and hitting time both agree well with the theoretical predictions. The numerically fitted results of classical optimal hitting efficiency $e_C$ and quantum optimal hitting efficiency $e_Q$ are presented in Table S4. The fitted results of classical hitting time $t_C$ and quantum hitting time $t_Q$ are presented in Table S5. The goodness of fitting is characterized by the coefficient of determination ($R^2$). From the fitting results, the optimal hitting occurs in a polynomial time both for CTQWs and CTRWs. For quantum hitting, the optimal hitting efficiency decays polynomially with the increase of $N$. In contrast, the scaling of classical optimal hitting efficiency can be expressed in exponential models, meaning the optimal hitting efficiency decays exponentially with $N$. These results demonstrate clear exponential speedups of CTQW hitting performance over classical scenarios on eRGTs and eCubes.

For eNets with layer depth $N$ up to 20 and eGrids with $N$ up to 36, comparisons between the CTQWs and CTRWs in the optimal hitting efficiency and hitting time are presented in experimental data and theoretical prediction in Table S3, Fig. S14 and Fig. S15. Overall, there is

good agreement between the experimental results and the theoretical predictions. The numerically fitted results of classical optimal hitting efficiency $e_C$ and quantum optimal hitting efficiency $e_Q$ are presented in Table S6, and the fitting of classical hitting time $t_C$ and quantum hitting time $t_Q$ are presented in Table S7. The goodness of fitting is characterized by the coefficient of determination ($R^2$). From the fitting results, the optimal hitting occurs in a linear time for CTQWs, while in a quadratic time for CTRWs. For quantum hitting, optimal hitting efficiency decays linearly with the increase of $N$. In contrast, the optimal hitting efficiency decays quadratically with $N$. These results show quadratic enhancement both in the hitting efficiency and hitting time of CTQW over classical scenarios on eNets and eGrids.

## 5. Details for quantum walk mixing

The most fundamental property of Markov chains is the truth that they converge to a stationary distribution independent of the initial state. However, quantum walks never "mix", that is, converge to any stationary state, due to the property that evolutions are unitary and reversible. Yet we can obtain a natural concept of convergence in the quantum case by sampling from the graph in a long-time probability distribution.

After CTQW for time $T$, the time-averaged probability that the walker is localized at the vertex $w$ is

$$\overline{\mathrm{P}}_w(T) = \frac{1}{T}\int_0^T dt \mathrm{P}_w(t) = \frac{1}{T}\int_0^T dt \left|\langle w|\mathrm{e}^{-iAt}|\phi_0\rangle\right|^2. \tag{30}$$

We can rewrite the unitary transformation $\mathrm{e}^{-iHt}$ using spectrum decomposing as

$$\mathrm{e}^{-iAt} = \sum_{\lambda_p} \lambda_p |\lambda_p\rangle\langle\lambda_p|, \tag{31}$$

where $\lambda_p$ is the eigenvalue and $|\lambda_p\rangle$ the corresponding eigenvector of the adjacent matrix $A$. Then

$$\overline{P}_w(T) = \frac{1}{T}\int_0^T dt \sum_{p,q} e^{-i(\lambda_p - \lambda_q)t} \langle w|\lambda_p\rangle\langle\lambda_q|w\rangle\langle\lambda_p|\phi_0\rangle\langle\phi_0|\lambda_q\rangle$$

$$= \sum_{\lambda_p = \lambda_q} \langle w|\lambda_p\rangle\langle\lambda_q|w\rangle\langle\lambda_p|\phi_0\rangle\langle\phi_0|\lambda_q\rangle \quad (32)$$

$$+ \sum_{\lambda_q \neq \lambda_q} \langle w|\lambda_p\rangle\langle\lambda_q|w\rangle\langle\lambda_p|\phi_0\rangle\langle\phi_0|\lambda_q\rangle \frac{1 - e^{-i(\lambda_p - \lambda_q)T}}{i(\lambda_p - \lambda_q)T}.$$

Let

$$\Delta_{p,q} = \lambda_p - \lambda_q, \quad (33)$$

$$w_{p,q} = \langle w|\lambda_p\rangle\langle\lambda_q|w\rangle\langle\lambda_p|\phi_0\rangle\langle\phi_0|\lambda_q\rangle, \quad (34)$$

we have that

$$\overline{P}_w(T) = \sum_{\lambda_p = \lambda_q} w_{p,q} + \sum_{\lambda_p > \lambda_q} \frac{4\sin\frac{\Delta_{p,q}T}{2}}{\Delta_{p,q}T}(\cos\frac{\Delta_{p,q}T}{2}\mathrm{Re}[w_{p,q}] + \sin\frac{\Delta_{p,q}T}{2}\mathrm{Im}[w_{p,q}]). \quad (35)$$

The first term is unrelated to time $T$, while the second term converges to zero as $T \to \infty$. Thus as $T \to \infty$, the mixing distribution approaches its limiting mixing asymptotically

$$\Pi_{ave,w} = \overline{P}_w(T \to \infty) = \sum_{\lambda_p = \lambda_q} w_{p,q}. \quad (36)$$

We say that $T$ is an $\varepsilon$-average mixing time for a quantum walk if $\|\overline{P}(T) - \Pi_{ave}\| \leq \varepsilon$, where

$$\|A - B\| = \frac{1}{2}\sum_i |A_i - B_i| \quad (37)$$

denotes total variation distance. Note that the sum in (36) is only on pairs such that $\lambda_p = \lambda_q$, thus, the average mixing dynamics depends on all the pairs with equal eigenvalues, while all the spectral gaps $\Delta_{p,q}$ are of crucial importance during the mixing process and convergence time. A detailed and further discussion of the average mixing dynamics of CTQW can be found in [29].

In addition to the average mixing, instantaneous mixing has been reported in quantum walks on some special graphs, such as one-dimensional lines, cycles, two-dimensional eGrids, and hypercubes. We say that $T$ is an $\varepsilon$-instantaneous mixing time for a quantum walk if $\|P(T) - \Pi_{uni}\| \leq \varepsilon$, where $\Pi_{uni}$ denotes the uniform distribution on vertices.

We are surely not going to approximate the limiting distribution by infinitely dense sampling. The general solution is to sample uniformly or randomly at fixed rate. However, poor sampling strategies degrade the results. Next, we analyze the relationship between the sampling rate and sampling error.

Suppose we experimentally investigate the dynamics of CTQW on graph $G$ with initial state $|\phi_0\rangle$, and the evolution time is $T$. When we set the total sample number as $N_s$, the sample interval is $\tau = T/N_s$. We begin our analysis with the experimentally obtained average probability in the sum form (instead of integrals) as

$$\overline{P}_w^{exp}(T) = \frac{1}{N_s}\sum_{n=1}^{N_s}|\langle w|e^{-iAn\tau}|\phi_0\rangle|^2$$
$$= \frac{1}{N_s}\sum_{n=1}^{N_s}\sum_{p,q}e^{-i(\lambda_p-\lambda_q)n\tau}\langle w|\lambda_p\rangle\langle\lambda_q|w\rangle\langle\lambda_p|\phi_0\rangle\langle\phi_0|\lambda_q\rangle. \quad (38)$$

Following (33) and (34), we can rewrite the average probability as

$$\overline{P}_w^{exp}(T) = \Pi_w + \frac{1}{N_s}\sum_{\lambda_p\neq\lambda_q}\frac{1-e^{i\Delta_{p,q}N_s\tau}}{1-e^{i\Delta_{p,q}t}}w_{p,q}$$
$$= \Pi_w + \sum_{\lambda_p>\lambda_q}\frac{2\sin\frac{\Delta_{p,q}T}{2}}{N\sin\frac{\Delta_{p,q}T}{2N_s}}(\cos\frac{\Delta_{p,q}T}{2}\mathrm{Re}[w_{p,q}]+\sin\frac{\Delta_{p,q}T}{2}\mathrm{Im}[w_{p,q}]). \quad (39)$$

Compare (35) and (39), we define the approximating error between theoretical and experimental mixing probability as

$$\epsilon_w = \max_{p,q}\frac{\left(\overline{P}_w^{exp}(T)-\Pi_w\right)-\left(\overline{P}_w(T)-\Pi_w\right)}{\overline{P}_w(T)-\Pi_w} = \max_{p,q}\frac{\frac{\Delta_{p,q}T}{2N_s}}{\sin\frac{\Delta_{p,q}T}{2N_s}}-1 = \frac{\frac{\Delta_A T}{2N_s}}{\sin\frac{\Delta_A T}{2N_s}}-1, \quad (40)$$

where $\Delta_A = \max_{p,q}\Delta_{p,q}$ is the gap between the maximum and minimum eigenvalues. We can see the sampling error is closely related to the spectrum range of the adjacent matrix, and the approximating error is independent of the vertex $w$. Thus, let $N_A = \Delta_A T$, the overall approximating error is

$$\epsilon = \frac{\frac{N_A}{2N_s}}{\sin\frac{N_A}{2N_s}} - 1. \tag{41}$$

According to the monotonicity of the function $f(x) = \frac{x}{\sin x} - 1$, we obviously verify that more sample number $N$ lowers the approximating error $\varepsilon$. See more details in Fig. S4.

In most experimental demonstrations of quantum mixing dynamics, we mainly set the sample number as

$$N_s = N_A, \tag{42}$$

then we get a proper error level as

$$\epsilon \leq 4.29\%. \tag{43}$$

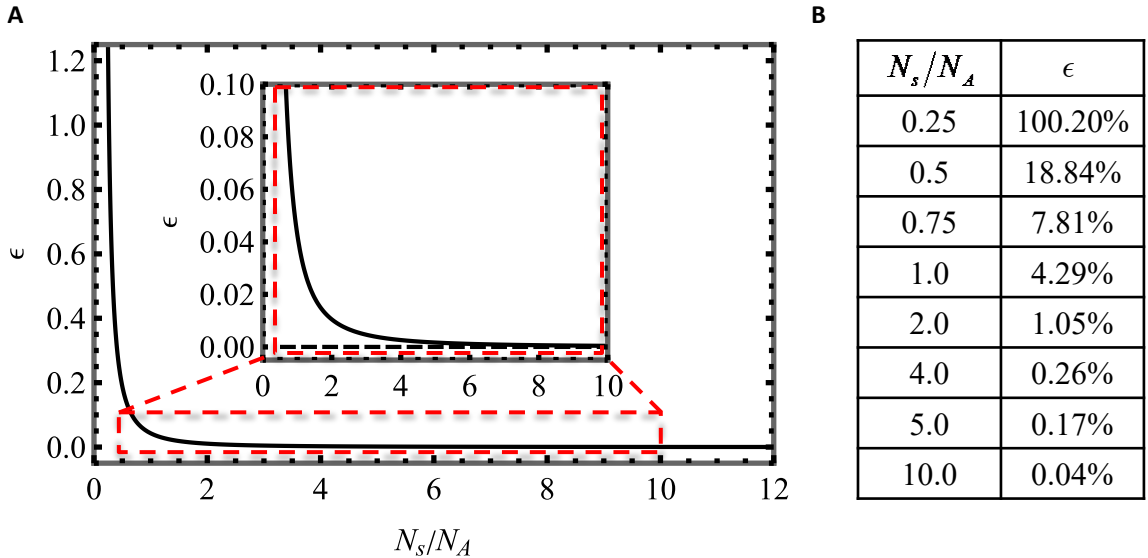

**Fig. S4 The overall approximating error with different sample numbers.** The function image of (41) are plotted in (A), and the error values with some special sample numbers are presented in table (B).

More experimental details in the simulations of mixing dynamics of CTQW on eNets and eGrids are listed in Table S8. For eNets with layer depth $N$ up to 20 and eGrids with $N$ up to 36, comparisons between the CTQWs and CTRWs in $\varepsilon$ – mixing time ($\varepsilon = 0.25$) are presented in experimental data and theoretical prediction in Table S9, Fig. S16, and Fig. S17 compare the

mixing time and evolution of total variance distance between the CTQWs and CTRWs. Overall, there is good agreement between the experimental results and the theoretical predictions. The numerically fitted results of classical mixing time $t_C$ and quantum mixing time $t_Q$ are presented in Table S10. The goodness of fitted results is characterized by the coefficient of determination ($R^2$). For classical mixing, the quadratic models obtain fairly accurate fitting, while for quantum mixing, linear models have sufficient precision. These results show quadratic speedups in the mixing time of CTQW over classical scenarios on eNets and eGrids.

## 6. Applications

### 6.1 Centrality measure

The quantum nature allows CTQW to propagate through networks faster than its classical counterpart, providing a potential efficient quantum algorithm for vertex centrality ranking in network analysis [30]. The core ideas root in the mixing dynamics of CTQW. When the quantum walker is initialized from the equal superposition over all vertices as

$$|\phi_0\rangle = \frac{1}{N}\sum_j |j\rangle, \qquad (44)$$

we can get a mixing distribution $\Pi_{ave}$ that correlates well with classical measures, such as the eigenvector centrality. We executed the CTQW centrality measure algorithm on eScale-free random networks by observing the average distribution from uniform superposition and after $\varepsilon$ – mixing time ($\varepsilon = 0.3$).

Extended figures present centrality measure results on other two eScale-free networks (Fig. S18). Comparisons between the CTQW centralities and eigenvector centralities and the mixing process on top-ranked vertices are presented. More experimental details of centrality measure are presented in Table S11.

## 6.2 Spatial search

Quantum walks provide a natural framework for the spatial search problem of finding marked items in an unsorted database of $N$ items, because the network can be used to model the locality of the database. It has been shown when tackling both single and multi-target search problems, the quantum walk-based search algorithm holds quadratic speedup over classical algorithms in almost all graphs [20]. Specifically, when searching $N_w$ marked vertices (denoted as $T = \{w_1, w_2, \ldots, w_{N_w}\}$) of network $G$, the walk evolution starts with a superposition state over $N_r$ randomly chosen vertices (denoted as $S = \{r_1, r_2, \ldots, r_{N_r}\}$, $S \cap T = \emptyset$), and the search Hamiltonian can be defined as

$$H_{search} = -\sum_{i=1}^{n_r}|r_i\rangle\langle r_i| - \sum_{j=1}^{n_w}|w_j\rangle\langle w_j| - \lambda A, \tag{45}$$

where $A$ is the adjacency matrix of the network, and $\lambda$ is the jump rate. Then the probability of finding one marked vertex is given by

$$P_w(t) = \sum_{j=1}^{n_w}\langle w_j|e^{itH_{search}}|\phi_S\rangle. \tag{46}$$

where $|\phi_S\rangle = \frac{1}{n_r}\sum_{j=1}^{n_r}|r_j\rangle$ is the initial superposition state. When $N_r \leq N_w \ll N$, $P_w$ can achieve O(1) with an evolution time $t = \mathrm{O}(\sqrt{N})$. By contrast, no classical algorithm can do better than exhaustive search, which takes of order O($N$) queries.

Until now, no physical experiment demonstrates the speedup performance of quantum walk-based search algorithm on random networks. By virtue of the large scale and full programmability of the system, we benchmarked the quantum walk algorithm on 1000 test cases with the network scale from 36 to 210. For each test case, we constructed a two-boson graph from a random Erdos-Renyi network $G(N, p)$ with $p = 0.5$, and searched for three targets from three initial vertices. Then we implemented the CTQW evolutions with the corresponding search

Hamiltonians (45), and measured the probabilities of achieving the target vertices at each time step. The search efficiency is defined as the proportion of probabilities on target vertices to probabilities on vertices excluding start vertices. We consider that optimal search occurs when the search efficiency has a deviance of no more than 5% from optimal efficiency. More experimental details of search tests are presented in Table S12. The fitting curve of the experimentally obtained optimal search time is

$$0.802556\sqrt{N} + 4.06895, \tag{47}$$

validating the quadratic advantage of CTQW-based search over classical scenarios.

### 6.3 Graph isomorphism

The graph isomorphism (GI) is a decision problem to determine whether two graphs are isomorphic. By isomorphic, we mean that the two graphs have identical structures by relabeling their vertices. GI has extensive applications in pattern recognition and computer vision.

It is still not clear whether GI is solvable in polynomial time or to be NP-complete, and therefore is thought to be in the NP-intermediate computational complexity class. Here, we utilize a QW-based algorithm [20] to solve the problem. Specifically, we construct the graph certificate $C_G$ for the given graph $G$, and distinguish the two graphs as non-isomorphic if the two obtained graph certificates are not equal. The graph certificate is a sorted list defined as

$$C_G = \text{sort}\left(\left\{P_i^{|i\rangle}(t), \text{for } i = 1, 2, \cdots, N\right\}\right) \tag{48}$$

where $P_i^{|i\rangle}(t)$ is the probability given by (8). To ensure the traversed symmetry, we need to iterate over the possible input vertex $i$. The pseudo code to calculate graph certificates is presented in Algorithm 2.

---

**Algorithm 2** Calculate $C_G$ [20]

---
**Input:** the adjacency matrix $A$ of graph $G$, evolution time $t$
1: $N \leftarrow |A|$             ▷ number of graph vertices
2: $\text{Prob} \leftarrow \{0, 0, \cdots, 0\}$        ▷ define Prob vector with length $N$
3: $l \leftarrow 0;$

33

4: **for** $i = 1:N$ **do**                  ▷ label of output and output vertex
5:   $\text{Prob}(i) \leftarrow P_i^{|i\rangle}(t)$;
6: **end**
7: **return** $C_G \leftarrow \text{sort}(\text{Prob})$

---

In Fig.3C, we give an instance of GI determinations of 210-vertex isomorphic graph pair and non-isomorphic graph pair, of which graphs are generated via two-boson CTQW on 20-vertex graphs. As stated in Algorithm 2, we prepare all the possible 210 input states, evolve under the graph adjacency matrix, and obtain the probability amplitude on the corresponding output vertex. To eliminate the fluctuations, we choose a sufficient long evolution time ($t_0 = 10^5$) and conduct the simulations for ten sample steps with timestep as 1. The distance between the two graphs $G_1$ and $G_2$ is evaluated by total variance distance $\|C_{G_1} - C_{G_2}\|$ as given by (37), and we set the average distance among ten timesteps as the evaluation metric. For the isomorphic graph pair, the average distance is close to 0 as $0.1013$, in contrast to the value of the non-isomorphic graph pair as high as $1.0122$. The average of fidelities in the experiments for graph certificates reaches $94.76 \pm 1.08\%$. More experimental results are presented in Table S13.

## 6.4 Exploring topological phases of HOTI

Here, we provide more details about the simulations of the bulk topology of two kinds of higher-order topological insulators (HOTIs): the 2D Su-Schrieffer-Heeger (SSH) model [32] and Benalcazar-Bernevig-Hughes (BBH) model [33] on *QW2020* system.

### 6.4.1 Simulating the topological phases of the 2D SSH model

The 2D SSH model describes spinless fermions hopping on a two-dimensional lattice with staggered hopping amplitudes, where the intracellular amplitudes are *v* and intercellular amplitudes are *w*, as shown in Fig. S5A. The 2D SSH model manifests topological phases when $|v/w| < 1$. We denote the Hamiltonian of the 2D SSH model as $H_{\text{S\_2D}}$. And we can write $H_{\text{S\_2D}}$ into the Kronecker sum of two 1D Hamiltonians $H_{\text{S\_1D}}$, which implies the 2D SSH model can

34

actually be reduced to one-dimensional systems, that is, the SSH model (as shown in Fig. S5B), on two spatial dimensions, respectively,

$$H_{S\_2D} = H_{S\_1D} \oplus H_{S\_1D} = I_x \otimes H^y_{S\_1D} + H^x_{S\_1D} \otimes I_y, \tag{49}$$

$$H^x_{S\_1D} = H^y_{S\_1D} = \sum_m v(a^\dagger_{m,A} a_{m,B} + h.c.) + \sum_{m'} w(a^\dagger_{m',B} a_{m'+1,A} + h.c.), \tag{50}$$

where $I_x$ and $I_y$ are identity operators, $a^\dagger_{m,\alpha}$ and $a_{m,\alpha}$ represent the creation and annihilation operators of spinless fermions at the vertex $\alpha$ ($\alpha$ = A, B) in the $m$-th unit cell of the SSH model.

*Maffei et al.* [34] have demonstrated that the topological properties of the SSH model, like winding number or Zak phase, can be revealed through the long-time averaged values of a bulk observable, mean chiral displacement (MCD). We denote the MCD as $\langle m(t) \rangle$ and define its long-time averaged values as $\langle m \rangle_{aver}$

$$\langle m(t) \rangle = \sum_m [m\Gamma_a |a_m(t)|^2 + m\Gamma_b |b_m(t)|^2], \tag{51}$$

$$\langle m \rangle_{aver} = \frac{1}{T} \int_0^T dt \langle m(t) \rangle = \frac{\mathcal{W}}{2}, \tag{52}$$

where $m$ is the position of each unit cell, $a_m(t)$ and $b_m(t)$ are the amplitudes that particles occupy vertices A and B in the $m$-th unit cell. $\Gamma_a$ and $\Gamma_b$ are the eigenvalues of the chiral operator. We take 1 for $\Gamma_a$ and -1 for $\Gamma_b$ here. It is proved that $\langle m \rangle_{aver}$ would be equal to $\mathcal{W}/2$, where $\mathcal{W}$ is the winding number of SSH model and proportional to Zak phase, when the particles have specific localized initial states, i.e., $|a_{m=0}(t)|^2 = 1$ and the time of measurement is sufficiently long.

Here, we heuristically extend the MCD to another pair of observables $\langle \vec{m}(t) \rangle = \langle m_x(t), m_y(t) \rangle$ and its long-time averaged values $\langle \vec{m} \rangle_{aver}$ for the 2D SSH model, as shown in (53) and (54), which naturally reveal the 2D Zak phases for the 2D SSH model [32]

$$\langle \vec{m}(t) \rangle = \langle m_x(t), m_y(t) \rangle$$
$$= \left\langle \sum_{\vec{r}} \sum_\alpha r_x \Gamma^{(x)}_\alpha |a_{\vec{r},\alpha}(t)|^2, \sum_{\vec{r}} \sum_\alpha r_y \Gamma^{(y)}_\alpha |a_{\vec{r},\alpha}(t)|^2 \right\rangle, \tag{53}$$

$$\langle \vec{m} \rangle_{aver} = \frac{1}{T} \int_0^T dt \langle \vec{m}(t) \rangle = \left\langle \frac{1}{T} \int_0^T dt m_x(t), \frac{1}{T} \int_0^T dt m_y(t) \right\rangle = \left( \frac{\mathcal{W}_x}{2}, \frac{\mathcal{W}_y}{2} \right), \tag{54}$$

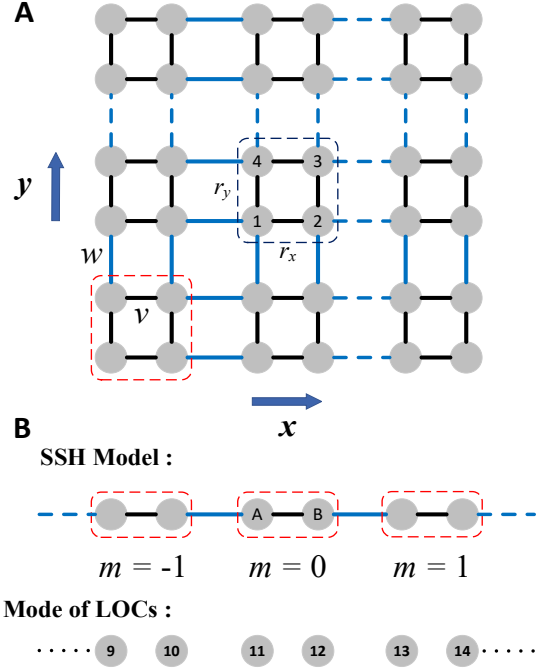

**Fig. S5 (A) Schematics of 2D SSH model with 10×10 unit cells.** The position of the unit cell in blue dashed box is $\vec{r} = (r_x, r_y)$, $r_x, r_y$ = -5, -4, …, 4, where $r_x$ and $r_y$ represent the positions along the direction $x$ and $y$, respectively. **(B) The mapping relationship between the vertices in the SSH model and spatial modes of linear optical circuits**.

where $m_x(t)$ and $m_y(t)$ are the MCDs of two SSH models, which are reduced from 2D SSH model on the $x$ and $y$ dimensions, $\vec{r} = (r_x, r_y)$ indicates the position of each unit cell in 2D SSH model, $a_{\vec{r},\alpha}(t)$ is the amplitude that particles occupy the vertex $\alpha = 1, 2, 3, 4$ in the unit cell at $\vec{r}$. $\Gamma_\alpha^{(x)}$ and $\Gamma_\alpha^{(y)}$ represent the eigenvalues of the chiral operators on dimensions $x$ and $y$, respectively. For $\Gamma_\alpha^{(x)}$, it should be 1 for $\alpha$ = 1, 4 and -1 for $\alpha$ = 2, 3, while for $\Gamma_\alpha^{(y)}$, it should be 1 for $\alpha$ = 1, 2 and -1 for $\alpha$ = 3, 4.

According to (49)-(52), we heuristically get that $\langle \vec{m} \rangle_{aver}$ would be equal to ($\mathcal{W}_x/2$, $\mathcal{W}_y/2$), which is equal to (1/2, 1/2) in topological phases and to (0, 0) for trivial phases, when particles are initially localized at the vertex 1 in the unit cell at $\vec{r}_0 = (0,0)$. Obviously, to obtain the extended MCD and its long-time averaged values, all we need to measure in experiments are the probability distributions of particles on all vertices of 2D SSH model. This means we can utilize

36

our CTQW-based system to simulate the topological phases and boundary dynamics of the 2D SSH model.

Since the Hamiltonian of the 2D SSH model can be written into the Kronecker sum of two SSH models Hamiltonians, we can construct it by implementing two fully distinguishable particles' CTQW on a one-dimensional lattice, which is the same as that described in SSH model. In this way, we should map the vertices in the SSH model to the spatial modes of the linear optical circuits on the photonic chip one by one. This implies that the maximum 2D SSH model we can simulate at the single-particle level would have $10\times10$ unit cells. In Fig. S5B, we give the mappings between the index of vertices in the SSH model and the number of spatial modes in linear optical circuits.

To simulate the topological phases of the 2D SSH model with $10\times10$ unit cells, firstly, we need to prepare the two-photon states where two photons both occupy the 11-th spatial mode of two linear optical circuits since these two modes correspond to the position $\vec{r}_0 = (0,0)$ in the 2D SSH model, as shown in Fig. S5B.

For CTQW, the related adjacency matrix of the graph should be

$$A_S = \begin{bmatrix} 0 & v & 0 & 0 & 0 & 0 & 0 & 0 & 0 & 0 \\ v & 0 & w & 0 & 0 & 0 & 0 & 0 & 0 & 0 \\ 0 & w & 0 & v & 0 & 0 & 0 & 0 & 0 & 0 \\ 0 & 0 & v & 0 & w & 0 & 0 & 0 & 0 & 0 \\ 0 & 0 & 0 & w & 0 & \ddots & 0 & 0 & 0 & 0 \\ 0 & 0 & 0 & 0 & \ddots & 0 & \ddots & 0 & 0 & 0 \\ 0 & 0 & 0 & 0 & 0 & \ddots & 0 & v & 0 & 0 \\ 0 & 0 & 0 & 0 & 0 & 0 & v & 0 & w & 0 \\ 0 & 0 & 0 & 0 & 0 & 0 & 0 & w & 0 & v \\ 0 & 0 & 0 & 0 & 0 & 0 & 0 & 0 & v & 0 \end{bmatrix}, \quad (55)$$

and the unitary transformations need to be set onto two linear optical circuits should be $e^{-iA_S t}$. Here, we choose the hopping amplitudes, $v = 1.0$, $w = 5.0$ for topological phases, and $v = 5.0$, $w = 1.0$ for trivial phases.

The probability distributions of single particle on vertices in the 2D SSH model after evolution time $t$ would be $\{P_{(x,y)}(t)|1 \leq x, y \leq 20\}$, as shown in (56). Then, we can use these to calculate the long-time averaged values of extended MCDs

$$P_{(x,y)}(t) = \left|\langle\psi_S(x,y)|(e^{-iA_St} \otimes e^{-iA_St})|\psi_S(0)\rangle\right|^2 \\ = \left|\langle 0|b_y a_x (e^{-iA_St} \otimes e^{-iA_St}) a_{11}^\dagger b_{11}^\dagger|0\rangle\right|^2. \tag{56}$$

### 6.4.2 Simulating the topological phases of the BBH model

*Benalcazar et al* [33] have proposed a tight-binding model with quantized quadrupole moment, that is, the BBH model, as shown in Fig. S6. In this model, the intracellular and intercellular hopping amplitudes are $\gamma$ and $\lambda$, respectively. The BBH model manifests topological phases when $|\gamma/\lambda| < 1$. The Hamiltonian of the BBH model without interactions between particles and disorders is

$$\boldsymbol{H}_B = \sum_{\langle i,j \rangle} t_{i,j} a_i^\dagger a_j, \tag{57}$$

where $\langle i, j \rangle$ represent two vertices connected directly by black or blue lines, as shown in Fig. S6.

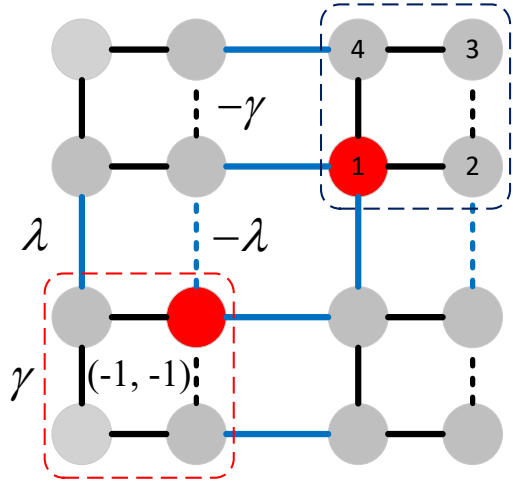

**Fig. S6 Schematics of BBH model with 2×2 unit cells.** The dash lines represent the hopping amplitudes with negative signs. The unit cells are also indexed by $\vec{r} = (r_x, r_y)$, $r_x$, $r_y$ = -1, 0. The vertices labeled red indicate the two-fermion initial states for simulating the bulk topology of the BBH model.

*Mizoguchi et al.* [35] have heuristically given a "good" observable, mean chiral quadrupole moment (MCQM) for probing the bulk topology of the BBH model with two-particle dynamics. The MCQM can also be seemed as an extended MCD

$$\langle m_q(t) \rangle = \sum_{\vec{r},\alpha} r_x r_y \Gamma_\alpha |a_{\vec{r},\alpha}(t)|^2, \tag{58}$$

where $\vec{r} = (r_x, r_y)$ indicates the position of each unit cell in the BBH model, $a_{\vec{r},\alpha}(t)$ is the amplitude that particles occupy the vertex $\alpha$ ($\alpha$ = 1, 2, 3, 4) in the unit cell at $\vec{r} = (r_x, r_y)$. $\Gamma_\alpha$ represent the eigenvalues of the chiral operators on dimensions $x$ and $y$, respectively. For $\Gamma_\alpha^{(x)}$, it should be 1 for $\alpha$ = 1, 3 and -1 for $\alpha$ = 2, 4. The long-time averaged values of MCQM $\langle m_q \rangle_{aver}$, as shown in (59), also is proved to be able to distinguish topological and trivial phases when two particles are localized at two diagonal vertices on the intercellular square at the center of the system, as shown in Fig. S6. Concretely, $\langle m_q \rangle_{aver}$ is equal to 1/2 in topological phases and 0 in trivial phases

$$\langle m_q \rangle_{aver} = \frac{1}{T} \int_0^T dt \langle m_q(t) \rangle. \tag{59}$$

Compared to the 2D SSH model, the BBH model can be seemed as a "true" two-dimensional system in some sense. At least, the Hamiltonian of the BBH model can't be constructed in the same way as we do for 2D SSH model. Here, we consider implementing two-fermion CTQWs on a 4×4 square lattice to simulate the topological phases of the BBH model with 2×2 unit cells. The mapping relation between the index of vertices in the BBH model $(\vec{r}, \alpha)$ and the number of spatial modes in linear optical circuits, $n$, satisfies $n = 4(r_x + Nr_y) + \alpha$, $N = 2$ here.

To simulate the topological phases of the BBH model with 2×2 unit cells, firstly, we need to prepare initial two-photon states as path-entangled states between the 3rd and 13th spatial modes of two linear optical circuits.

For CTQW, the related adjacency matrix of the graph should be

$$A_B = \begin{bmatrix}
0 & \gamma & 0 & \gamma & 0 & 0 & 0 & 0 & 0 & 0 & 0 & 0 & 0 & 0 & 0 & 0 \\
\gamma & 0 & -\gamma & 0 & \lambda & 0 & 0 & 0 & 0 & 0 & 0 & 0 & 0 & 0 & 0 & 0 \\
0 & -\gamma & 0 & \gamma & 0 & 0 & 0 & \lambda & 0 & -\lambda & 0 & 0 & 0 & 0 & 0 & 0 \\
\gamma & 0 & \gamma & 0 & 0 & 0 & 0 & 0 & \lambda & 0 & 0 & 0 & 0 & 0 & 0 & 0 \\
0 & \lambda & 0 & 0 & 0 & \gamma & 0 & \gamma & 0 & 0 & 0 & 0 & 0 & 0 & 0 & 0 \\
0 & 0 & 0 & 0 & \gamma & 0 & -\gamma & 0 & 0 & 0 & 0 & 0 & 0 & 0 & 0 & 0 \\
0 & 0 & 0 & 0 & 0 & -\gamma & 0 & \gamma & 0 & 0 & 0 & 0 & -\lambda & 0 & 0 & 0 \\
0 & 0 & \lambda & 0 & \gamma & 0 & \gamma & 0 & 0 & 0 & 0 & 0 & \lambda & 0 & 0 & 0 \\
0 & 0 & 0 & \lambda & 0 & 0 & 0 & 0 & 0 & \gamma & 0 & \gamma & 0 & 0 & 0 & 0 \\
0 & 0 & -\lambda & 0 & 0 & 0 & 0 & 0 & \gamma & 0 & -\gamma & 0 & \lambda & 0 & 0 & 0 \\
0 & 0 & 0 & 0 & 0 & 0 & 0 & 0 & 0 & -\gamma & 0 & \gamma & 0 & 0 & 0 & \lambda \\
0 & 0 & 0 & 0 & 0 & 0 & 0 & 0 & \gamma & 0 & \gamma & 0 & 0 & 0 & 0 & 0 \\
0 & 0 & 0 & 0 & 0 & 0 & 0 & \lambda & 0 & \lambda & 0 & 0 & 0 & \gamma & 0 & \gamma \\
0 & 0 & 0 & 0 & 0 & 0 & -\lambda & 0 & 0 & 0 & 0 & 0 & \gamma & 0 & -\gamma & 0 \\
0 & 0 & 0 & 0 & 0 & 0 & 0 & 0 & 0 & 0 & 0 & 0 & 0 & -\gamma & 0 & \gamma \\
0 & 0 & 0 & 0 & 0 & 0 & 0 & 0 & 0 & \lambda & 0 & \gamma & 0 & \gamma & 0 & 0
\end{bmatrix}, \quad (60)$$

and the unitary transformations need to be set onto two linear optical circuits should be $e^{-iA_S t}$. Here, we choose the hopping amplitudes, $\gamma = 0.1$, $\lambda = 4.0$ for topological phases, and $\gamma = 4.0$, $\lambda = 0.1$ for trivial phases.

The probability distributions of two particles on vertices in the BBH model after evolution time $t$ would be $\{P_{(i,j)}(t)|1 \leq i < j \leq 16\}$, as shown in (61). Then, we can use these to calculate the long-time averaged values of MCQM

$$\begin{aligned}
P_{(i,j)}(t) &= \left|\langle \psi_B(i,j)|(e^{-iA_B t} \otimes e^{-iA_B t})|\psi_B(0)\rangle\right|^2 \\
&= \left|\frac{1}{2}\langle 0|(b_i a_j - b_j a_i)\,(e^{-iA_B t} \otimes e^{-iA_B t})\,(a_3^\dagger b_{13}^\dagger - a_{13}^\dagger b_3^\dagger)|0\rangle\right|^2.
\end{aligned} \quad (61)$$

### 6.4.3 Extended data for **topological** phases simulation

We present the experimental details of simulating the topological properties of two HOTIs in Fig. S19, Fig. S20, Fig. S21, and Table S14, which show that we have implemented the simulations of the bulk topology of these models with high fidelity and high precision.

## 7. Extended data, figures, and tables

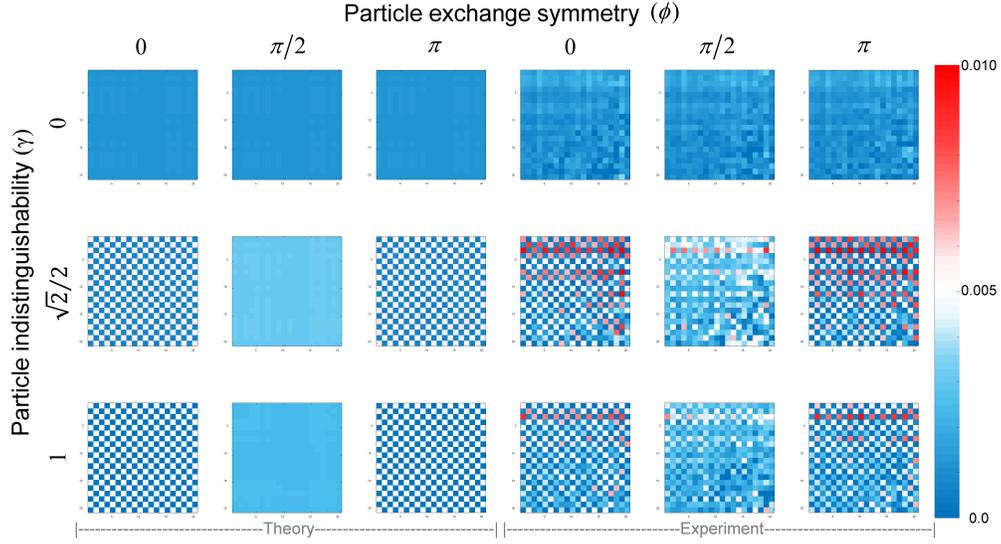

**Fig. S7 Experimental simulation of CTQW of two correlated particles on a 20-vertex circle with tunable indistinguishability and exchange symmetry at the evolution time π/8.** The average of fidelities between experimental and theoretical results is 95.62±1.13%.

In Fig. S3, we elaborate the theoretical and experimental quantum interference statistics diagrams at the evolution time $\pi/8$. The results show that particle correlations change substantially with particle properties, even in the same unitary evolution as

$$U = U_{\text{CTQW}} \cdot U_{\text{prep}} = e^{-itA_{C20}} \cdot U_{\text{prep}}, \tag{62}$$

where $A_{C20}$ is the adjacency matrix of the 20-vertex circle, and we apply $U_{\text{prep}}$ as

$$\begin{bmatrix}
\sqrt{\frac{1}{10}} & 0 & -\sqrt{\frac{1}{90}} & 0 & -\sqrt{\frac{1}{72}} & 0 & -\sqrt{\frac{1}{56}} & 0 & -\sqrt{\frac{1}{42}} & 0 & -\sqrt{\frac{1}{30}} & 0 & -\sqrt{\frac{1}{20}} & 0 & -\sqrt{\frac{1}{12}} & 0 & -\sqrt{\frac{1}{6}} & 0 & -\sqrt{\frac{1}{2}} & 0 \\
0 & \sqrt{\frac{1}{10}} & 0 & -\sqrt{\frac{1}{90}} & 0 & -\sqrt{\frac{1}{72}} & 0 & -\sqrt{\frac{1}{56}} & 0 & -\sqrt{\frac{1}{42}} & 0 & -\sqrt{\frac{1}{30}} & 0 & -\sqrt{\frac{1}{20}} & 0 & -\sqrt{\frac{1}{12}} & 0 & -\sqrt{\frac{1}{6}} & 0 & -\sqrt{\frac{1}{2}} \\
\sqrt{\frac{1}{10}} & 0 & \sqrt{\frac{9}{10}} & 0 & 0 & 0 & 0 & 0 & 0 & 0 & 0 & 0 & 0 & 0 & 0 & 0 & 0 & 0 & 0 & 0 \\
0 & \sqrt{\frac{1}{10}} & 0 & \sqrt{\frac{9}{10}} & 0 & 0 & 0 & 0 & 0 & 0 & 0 & 0 & 0 & 0 & 0 & 0 & 0 & 0 & 0 & 0 \\
\sqrt{\frac{1}{10}} & 0 & -\sqrt{\frac{1}{90}} & 0 & \sqrt{\frac{8}{9}} & 0 & 0 & 0 & 0 & 0 & 0 & 0 & 0 & 0 & 0 & 0 & 0 & 0 & 0 & 0 \\
0 & \sqrt{\frac{1}{10}} & 0 & -\sqrt{\frac{1}{90}} & 0 & \sqrt{\frac{8}{9}} & 0 & 0 & 0 & 0 & 0 & 0 & 0 & 0 & 0 & 0 & 0 & 0 & 0 & 0 \\
\sqrt{\frac{1}{10}} & 0 & -\sqrt{\frac{1}{90}} & 0 & -\sqrt{\frac{1}{72}} & 0 & \sqrt{\frac{7}{8}} & 0 & 0 & 0 & 0 & 0 & 0 & 0 & 0 & 0 & 0 & 0 & 0 & 0 \\
0 & \sqrt{\frac{1}{10}} & 0 & -\sqrt{\frac{1}{90}} & 0 & -\sqrt{\frac{1}{72}} & 0 & \sqrt{\frac{7}{8}} & 0 & 0 & 0 & 0 & 0 & 0 & 0 & 0 & 0 & 0 & 0 & 0 \\
\sqrt{\frac{1}{10}} & 0 & -\sqrt{\frac{1}{90}} & 0 & -\sqrt{\frac{1}{72}} & 0 & -\sqrt{\frac{1}{56}} & 0 & \sqrt{\frac{6}{7}} & 0 & 0 & 0 & 0 & 0 & 0 & 0 & 0 & 0 & 0 & 0 \\
0 & \sqrt{\frac{1}{10}} & 0 & -\sqrt{\frac{1}{90}} & 0 & -\sqrt{\frac{1}{72}} & 0 & -\sqrt{\frac{1}{56}} & 0 & \sqrt{\frac{6}{7}} & 0 & 0 & 0 & 0 & 0 & 0 & 0 & 0 & 0 & 0 \\
\sqrt{\frac{1}{10}} & 0 & -\sqrt{\frac{1}{90}} & 0 & -\sqrt{\frac{1}{72}} & 0 & -\sqrt{\frac{1}{56}} & 0 & -\sqrt{\frac{1}{42}} & 0 & \sqrt{\frac{5}{6}} & 0 & 0 & 0 & 0 & 0 & 0 & 0 & 0 & 0 \\
0 & \sqrt{\frac{1}{10}} & 0 & -\sqrt{\frac{1}{90}} & 0 & -\sqrt{\frac{1}{72}} & 0 & -\sqrt{\frac{1}{56}} & 0 & -\sqrt{\frac{1}{42}} & 0 & \sqrt{\frac{5}{6}} & 0 & 0 & 0 & 0 & 0 & 0 & 0 & 0 \\
\sqrt{\frac{1}{10}} & 0 & -\sqrt{\frac{1}{90}} & 0 & -\sqrt{\frac{1}{72}} & 0 & -\sqrt{\frac{1}{56}} & 0 & -\sqrt{\frac{1}{42}} & 0 & -\sqrt{\frac{1}{30}} & 0 & \sqrt{\frac{4}{5}} & 0 & 0 & 0 & 0 & 0 & 0 & 0 \\
0 & \sqrt{\frac{1}{10}} & 0 & -\sqrt{\frac{1}{90}} & 0 & -\sqrt{\frac{1}{72}} & 0 & -\sqrt{\frac{1}{56}} & 0 & -\sqrt{\frac{1}{42}} & 0 & -\sqrt{\frac{1}{30}} & 0 & \sqrt{\frac{4}{5}} & 0 & 0 & 0 & 0 & 0 & 0 \\
\sqrt{\frac{1}{10}} & 0 & -\sqrt{\frac{1}{90}} & 0 & -\sqrt{\frac{1}{72}} & 0 & -\sqrt{\frac{1}{56}} & 0 & -\sqrt{\frac{1}{42}} & 0 & -\sqrt{\frac{1}{30}} & 0 & -\sqrt{\frac{1}{20}} & 0 & \sqrt{\frac{3}{4}} & 0 & 0 & 0 & 0 & 0 \\
0 & \sqrt{\frac{1}{10}} & 0 & -\sqrt{\frac{1}{90}} & 0 & -\sqrt{\frac{1}{72}} & 0 & -\sqrt{\frac{1}{56}} & 0 & -\sqrt{\frac{1}{42}} & 0 & -\sqrt{\frac{1}{30}} & 0 & -\sqrt{\frac{1}{20}} & 0 & \sqrt{\frac{3}{4}} & 0 & 0 & 0 & 0 \\
\sqrt{\frac{1}{10}} & 0 & -\sqrt{\frac{1}{90}} & 0 & -\sqrt{\frac{1}{72}} & 0 & -\sqrt{\frac{1}{56}} & 0 & -\sqrt{\frac{1}{42}} & 0 & -\sqrt{\frac{1}{30}} & 0 & -\sqrt{\frac{1}{20}} & 0 & -\sqrt{\frac{1}{12}} & 0 & \sqrt{\frac{2}{3}} & 0 & 0 & 0 \\
0 & \sqrt{\frac{1}{10}} & 0 & -\sqrt{\frac{1}{90}} & 0 & -\sqrt{\frac{1}{72}} & 0 & -\sqrt{\frac{1}{56}} & 0 & -\sqrt{\frac{1}{42}} & 0 & -\sqrt{\frac{1}{30}} & 0 & -\sqrt{\frac{1}{20}} & 0 & -\sqrt{\frac{1}{12}} & 0 & \sqrt{\frac{2}{3}} & 0 & 0 \\
\sqrt{\frac{1}{10}} & 0 & -\sqrt{\frac{1}{90}} & 0 & -\sqrt{\frac{1}{72}} & 0 & -\sqrt{\frac{1}{56}} & 0 & -\sqrt{\frac{1}{42}} & 0 & -\sqrt{\frac{1}{30}} & 0 & -\sqrt{\frac{1}{20}} & 0 & -\sqrt{\frac{1}{12}} & 0 & -\sqrt{\frac{1}{6}} & 0 & \sqrt{\frac{1}{2}} & 0 \\
0 & \sqrt{\frac{1}{10}} & 0 & -\sqrt{\frac{1}{90}} & 0 & -\sqrt{\frac{1}{72}} & 0 & -\sqrt{\frac{1}{56}} & 0 & -\sqrt{\frac{1}{42}} & 0 & -\sqrt{\frac{1}{30}} & 0 & -\sqrt{\frac{1}{20}} & 0 & -\sqrt{\frac{1}{12}} & 0 & -\sqrt{\frac{1}{6}} & 0 & \sqrt{\frac{1}{2}}
\end{bmatrix}$$

onto the fixed input state to prepare the initial state of the two particles.

As indistinguishability gradually varies from 0 to 1, the particle correlations are passing from being influenced by totally classical effects $(\gamma = 0)$ to the stage where quantum interference becomes more dominant $(\gamma = 1/\sqrt{2})$, and ultimately, quantum interference completely determines $(\gamma = 1)$. As particle exchange symmetry alters, the two-particle interference statistics range from the Bose-Einstein statistics $(\phi = 0)$, to the intermediate fractional statistics $(\phi = \pi/2)$, and finally to Fermi-Dirac statistics $(\phi = \pi)$. Obviously, when two particles are fully indistinguishable, diagonal terms imply the anti-symmetry associated with fermions leads to the Pauli exclusion principle when $\phi = \pi$, while the symmetry associated with bosons enables bunching when $\phi = 0$. However, when indistinguishability weakens as $\gamma < 1$,

though diagonal terms with $\phi = \pi$ are still suppressed, the Pauli exclusion principle ceases to be effective.

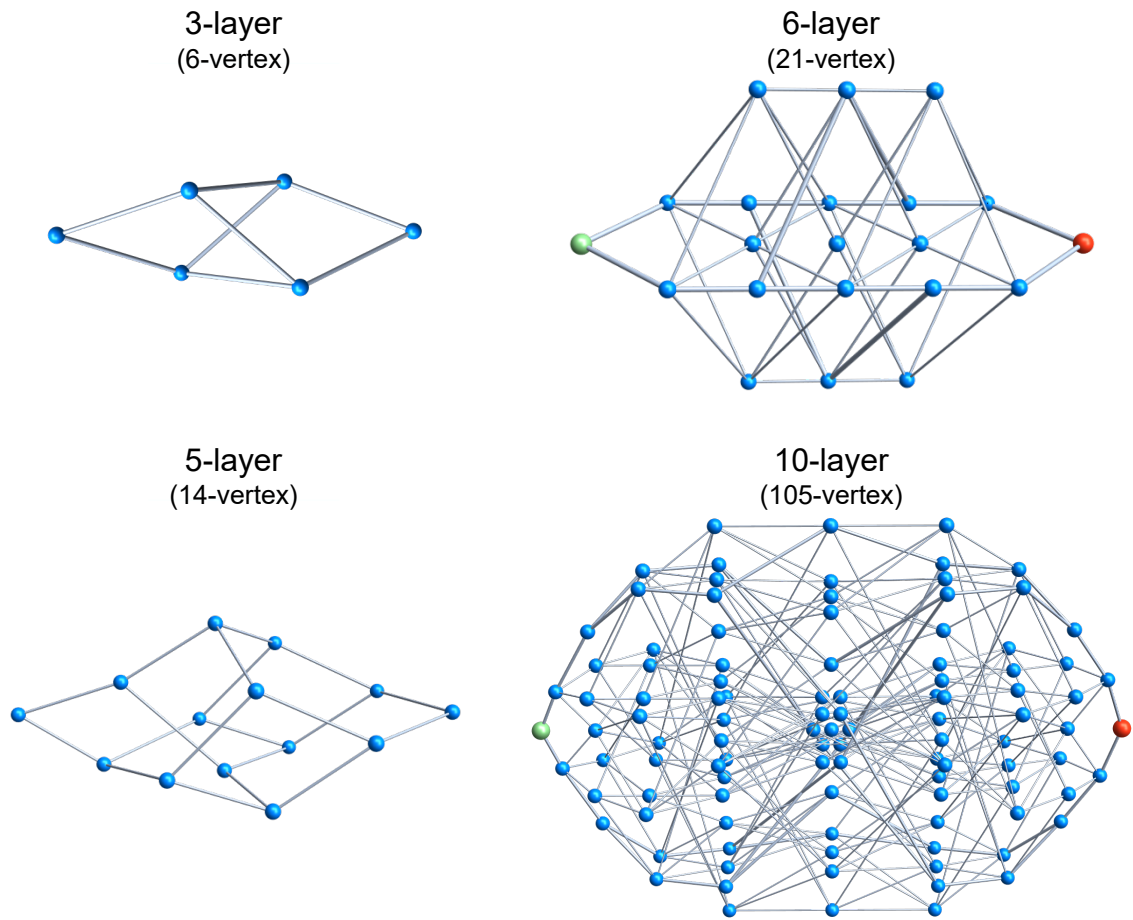

**Fig. S8** The 6- and 10-layer eRGTs generated via two-boson CTQW on 3-, and 5-layered random glued binary trees (RGTs), respectively. The entrance vertices in hitting experiments are colored in green, while the exits are red.

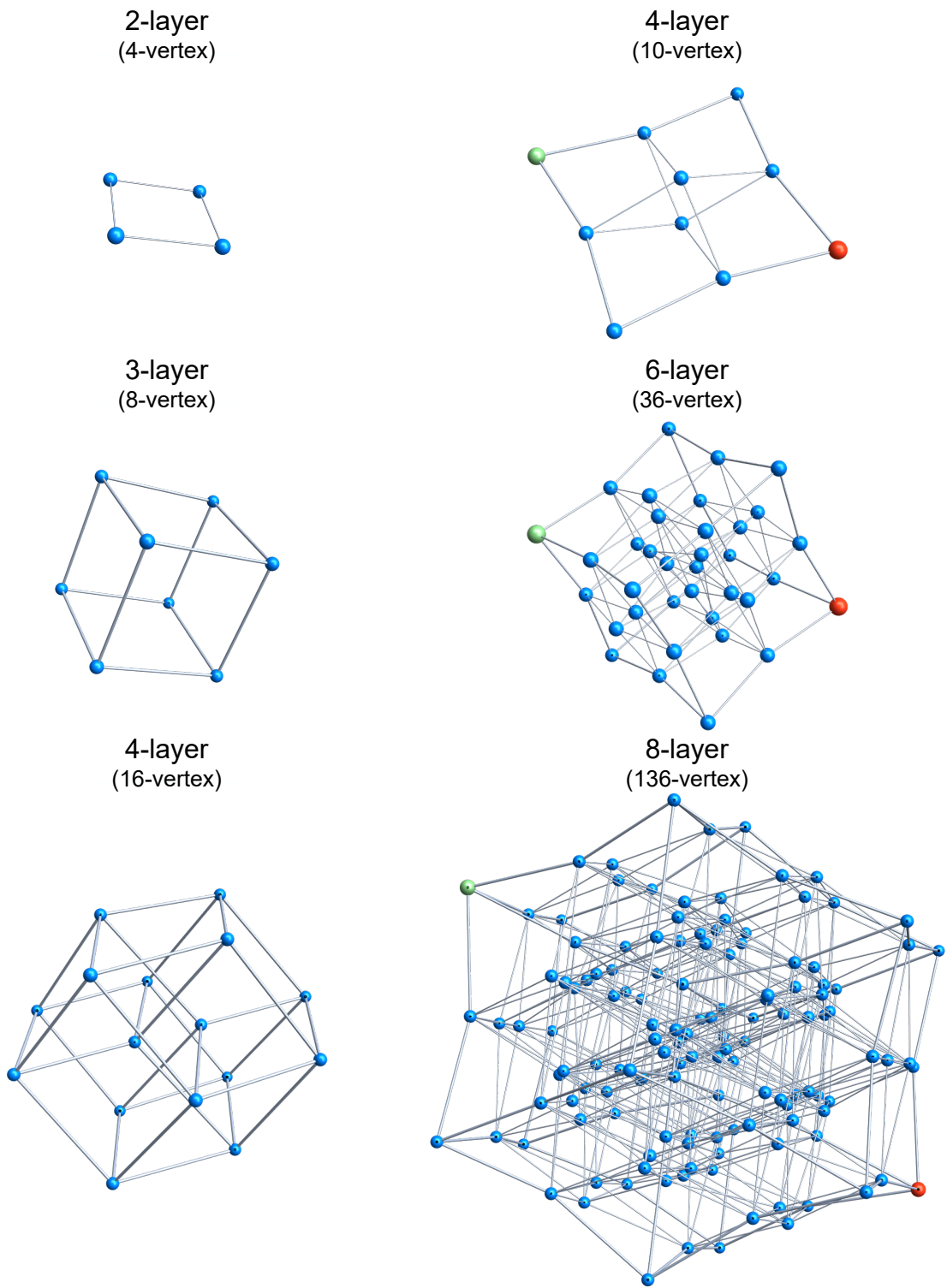

**Fig. S9** The 6-, 8-, and 10-layer eCubes generated via two-boson CTQW on 2-, 3-, and 4-layered hypercubes, respectively. The entrance vertices in hitting and mixing experiments are colored in green, while the exits (only in hitting experiments) are red.

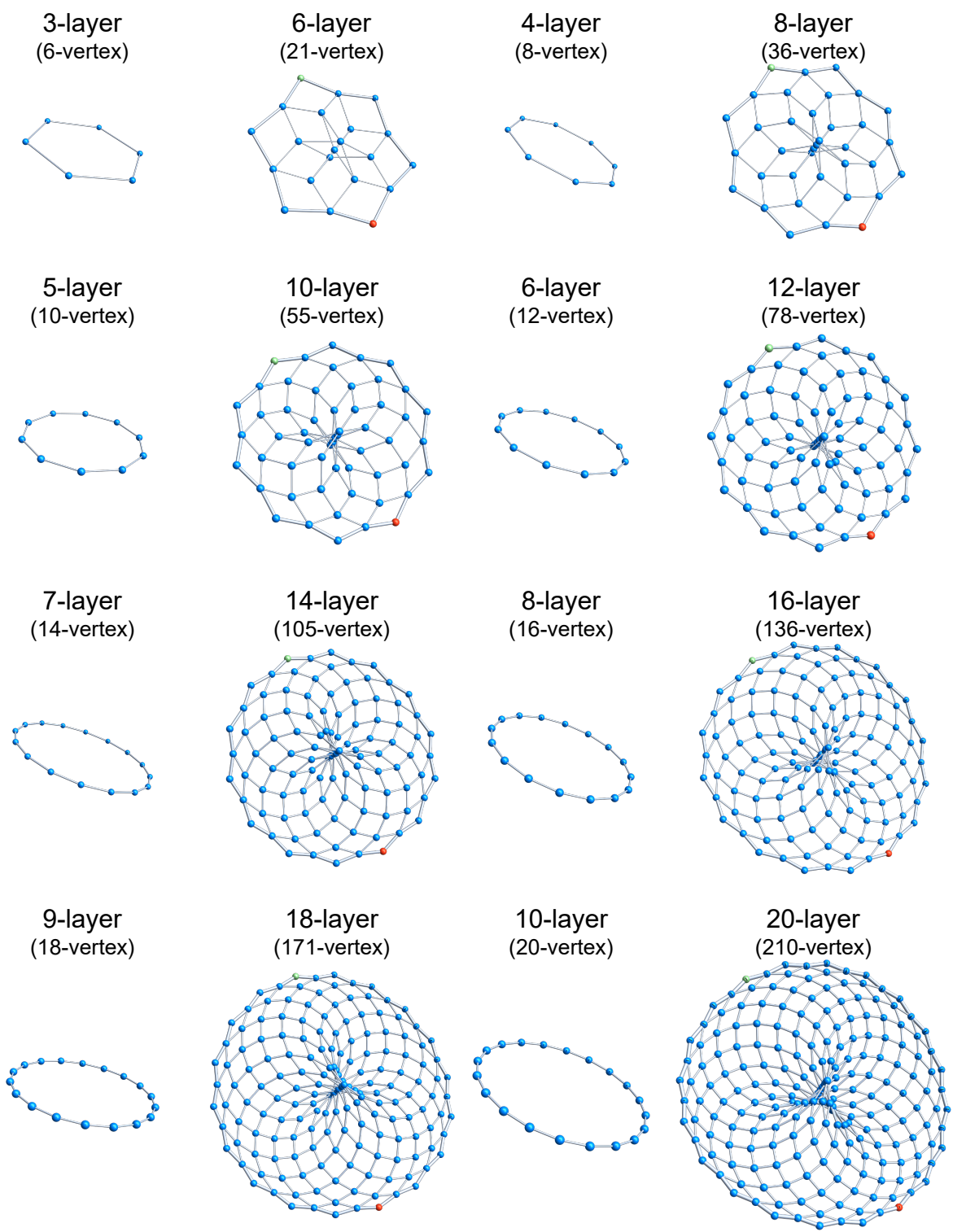

**Fig. S10** The 6-layer to 20-layer eNets generated via two-boson CTQW on 3-layer to 10-layer circles, respectively. The entrance vertices in hitting and mixing experiments are colored in green, while the exits (only in hitting experiments) are red.

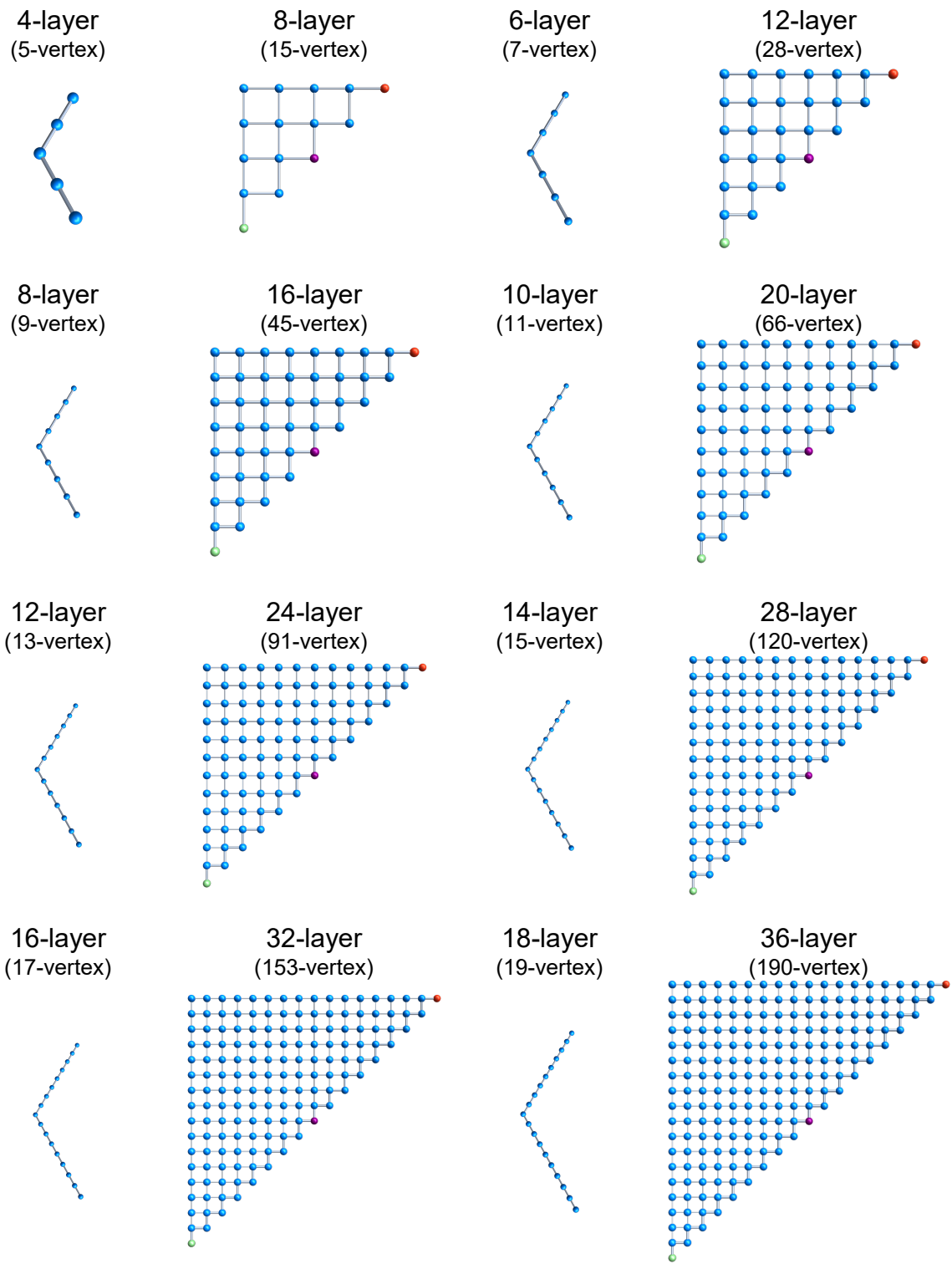

**Fig. S11** The 8-layer to 36-layer eGrids generated via two-boson CTQW on 4-layer to 18-layer lines, respectively. The entrance vertices in hitting and mixing experiments are colored in green and purple, respectively, while the exits (only in hitting experiments) are red.

46

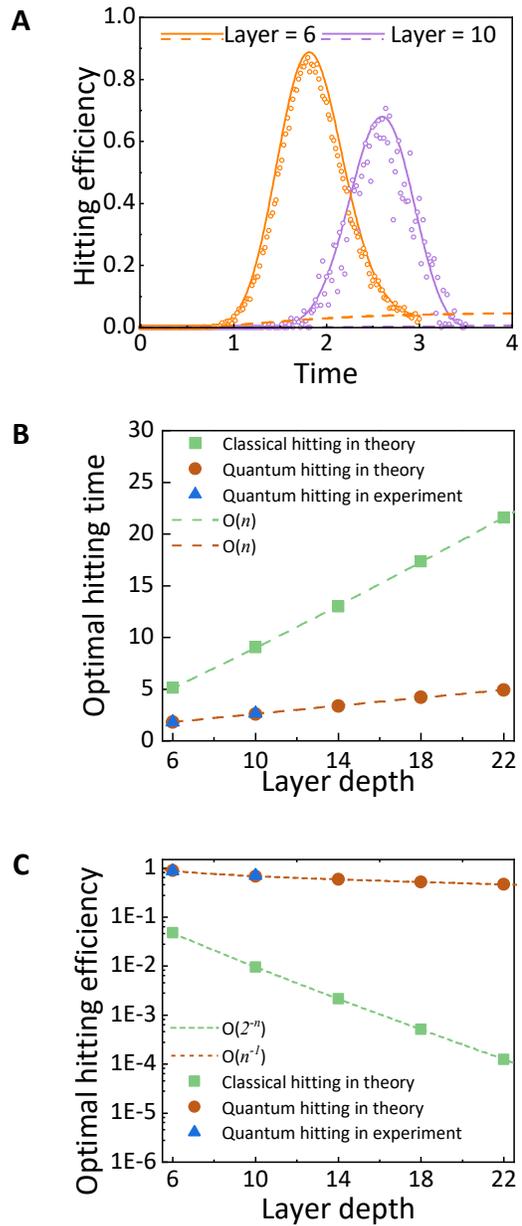

**Fig. S12 Exponentially fast hitting on eRGTs.** (A) The time-evolved hitting efficiency in quantum hitting (solid line) and classical hitting (dashed-dot line). Comparison in the (B) optimal hitting time and (C) optimal hitting efficiency between quantum hitting and classical hitting on generalized glued trees of different layer depths. Quantum walk and classical walk both achieve the optimal hitting in a time that is polynomial in layer depth, while the classical hitting efficiency must be exponentially smaller than the quantum counterpart.

47

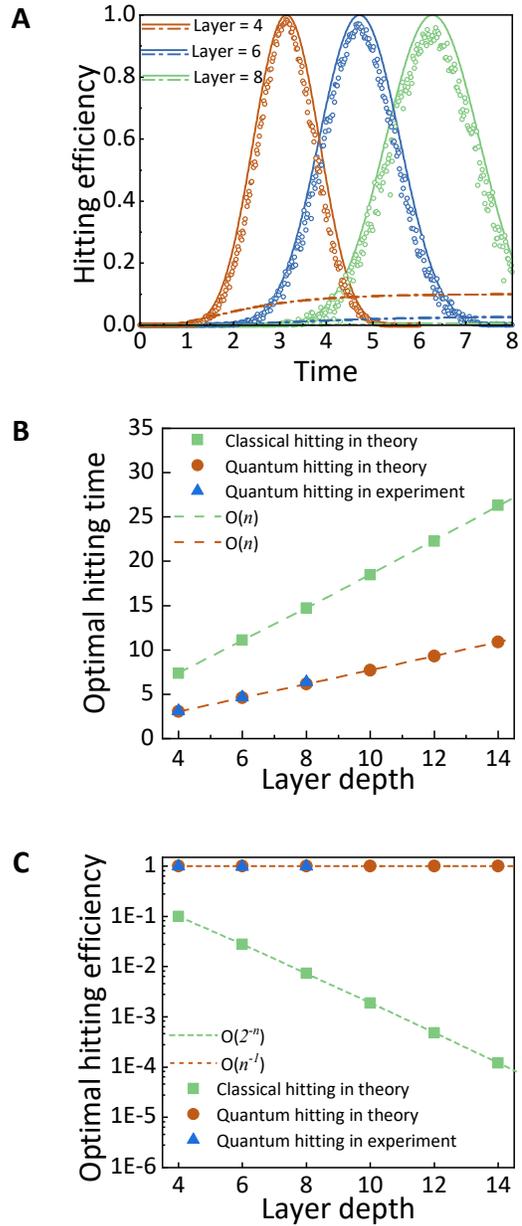

**Fig. S13 Exponentially fast hitting on eCubes.** (A) The time-evolved hitting efficiency in quantum hitting (solid line) and classical hitting (dashed-dot line). Comparison in the optimal hitting time (B) and optimal hitting efficiency (C) between quantum walk and classical random walk on eCubes of different layer depths. Quantum walk and classical walk both achieve the optimal hitting in a time that is polynomial in layer depth, while the classical hitting efficiency must be exponentially smaller than the quantum counterpart.

48

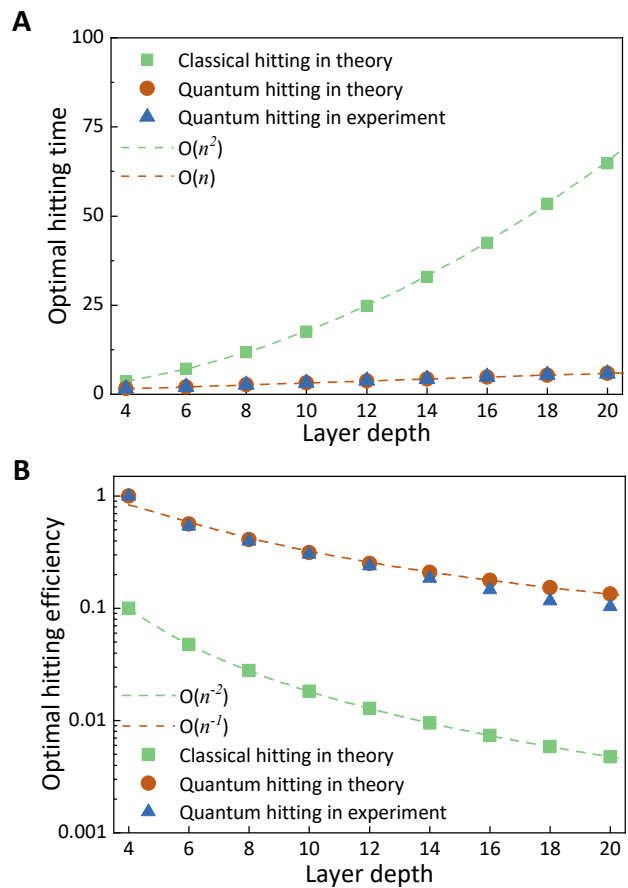

**Fig. S14 Quadratically fast hitting on eNets.** Comparison in the optimal hitting time (A) and optimal hitting efficiency (B) between quantum hitting and classical hitting on eNets of different layer depths.

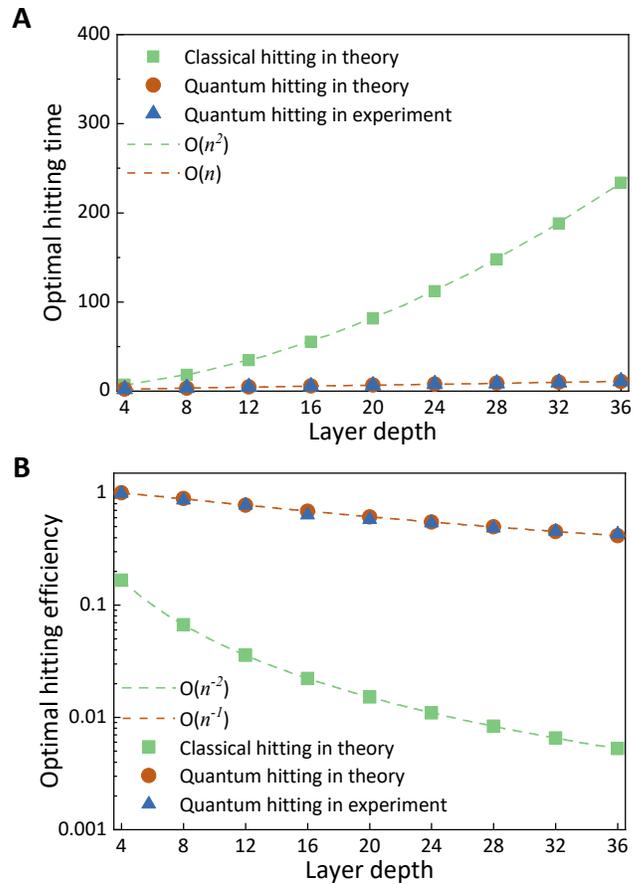

**Fig. S15 Quadratically fast hitting on eGrids.** Comparison in the optimal hitting time (A) and optimal hitting efficiency (B) between quantum hitting and classical hitting on eGrids of different layer depths.

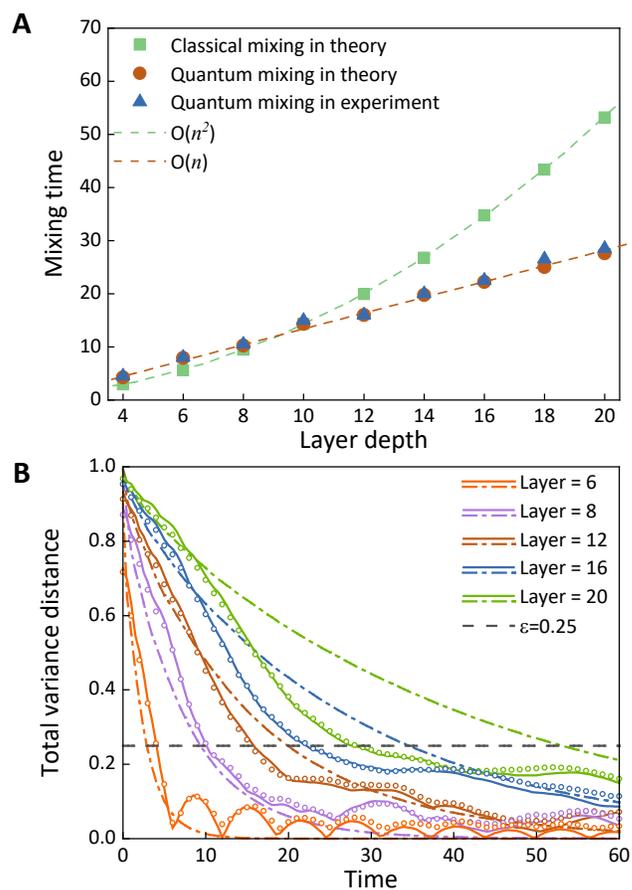

**Fig. S16 Quadratically fast mixing on eNets.** Comparison in the (A) $\varepsilon$–mixing time ($\varepsilon = 0.25$) and (B) total variance distance between quantum average mixing and classical mixing. In (B), solid and dashed lines show the theoretically predicted evolution of total variance distance from the mixing distribution of quantum and classical walk, respectively. Dots present the experimentally obtained results.

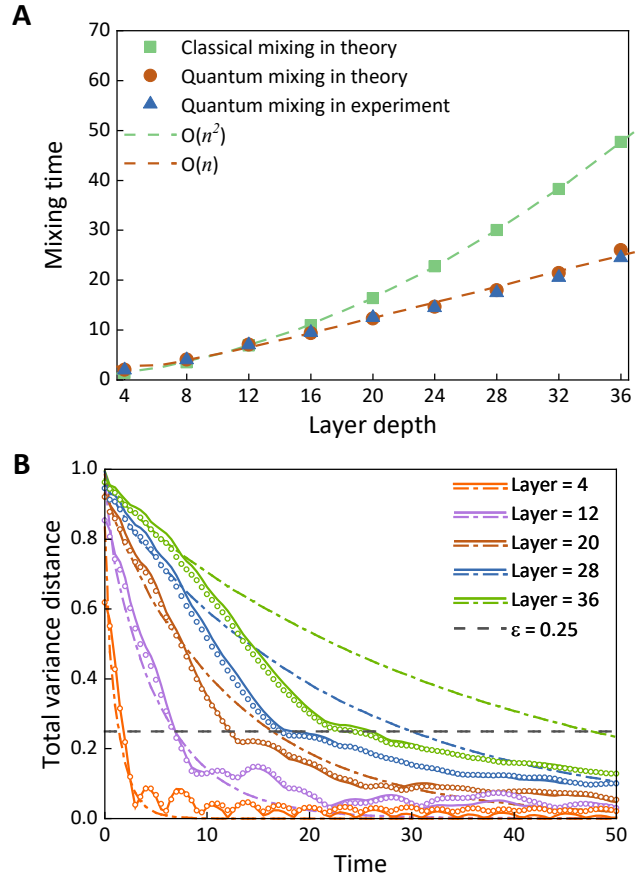

**Fig. S17 Quadratically fast mixing on eGrids.** Comparison in the (A) $\varepsilon$ – mixing time ($\varepsilon = 0.25$) and (B) total variance distance between quantum average mixing and classical mixing. In (B), solid and dashed lines show the theoretically predicted evolution of total variance distance from the mixing distribution of quantum and classical walk, respectively. Dots present the experimentally obtained results.

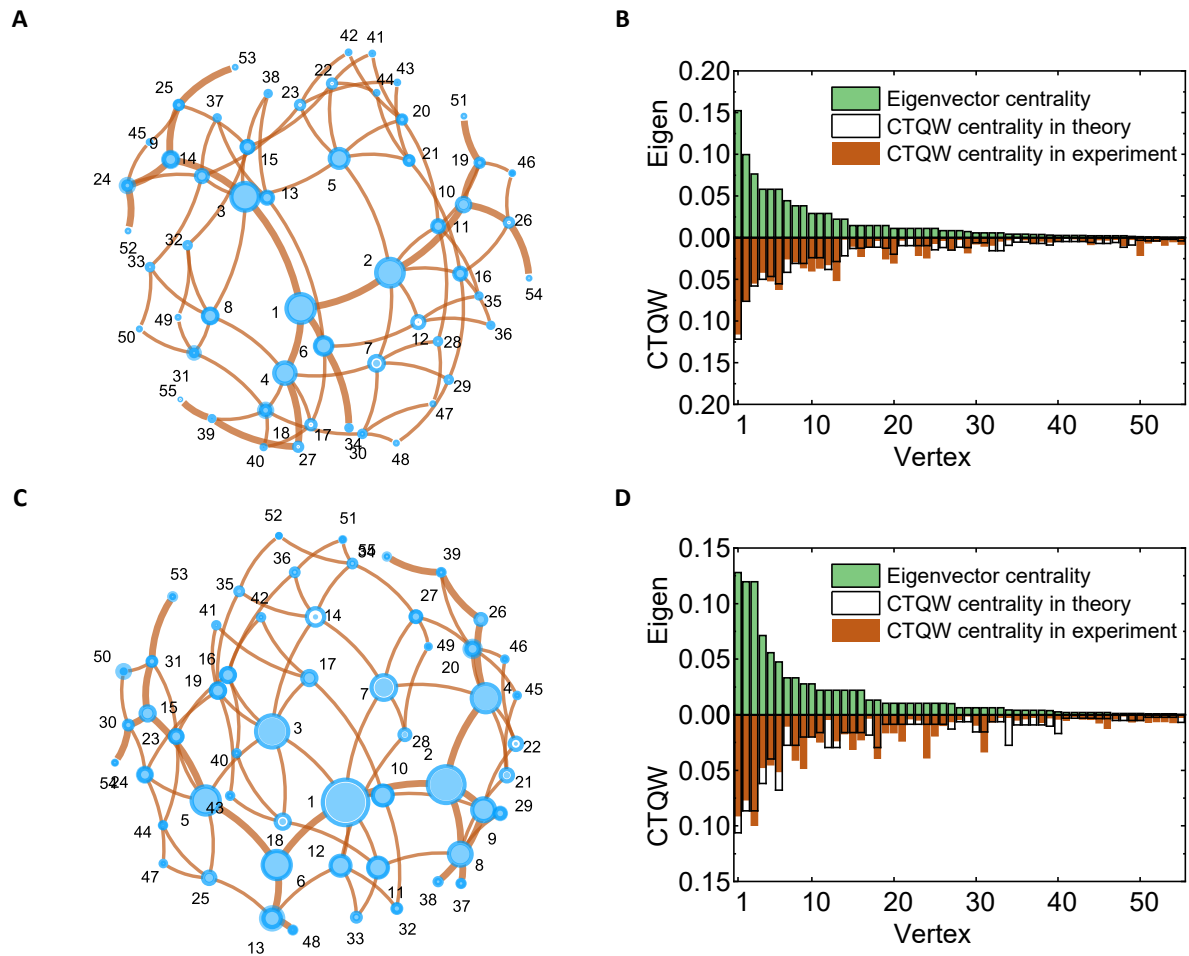

**Fig. S18 Centrality measure of two 55-vertex eScale-free networks based on CTQW mixing dynamics.** Two 55-vertex scale-free random networks are shown in (A) and (C), with vertex size indicating the vertex centrality. The empty circles represent the theoretical eigenvector centrality, with experimentally determined CTQW centrality value overlaid. (B) and (C) present the comparison between eigenvector centrality and CTQW centrality. The similarity of eigenvector centrality and experimentally-obtained CTQW centrality for the two networks are 95.09% and 95.01%, respectively. All centrality measures strongly agree on the top-ranked vertices, with slight variations or the lower-ranked vertices.

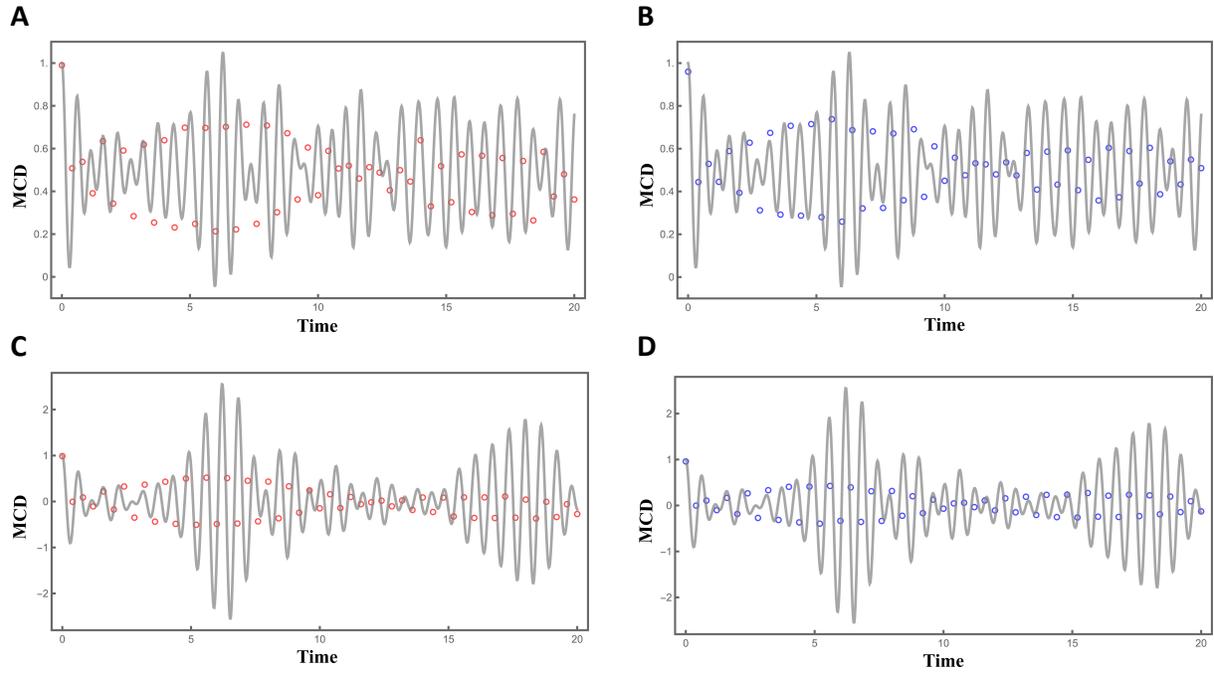

**Fig. S19 Instantaneous values of extended MCDs of 2D SSH model.** (A) and (B) show the instantaneous values of extended MCDs along the dimensions $x$ and $y$ in topological phases, respectively. (C) and (D) show the corresponding results in trivial phases. The gray lines represent theoretical values and red (or blue) circles are experimental values.

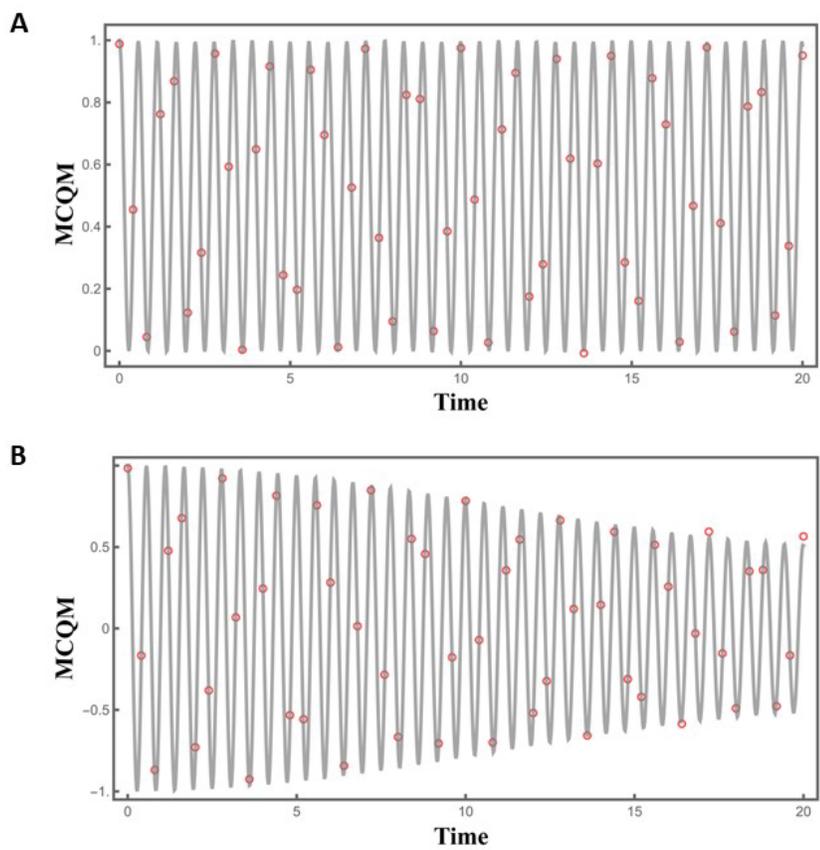

**Fig. S20 Instantaneous values of MCQMs of BBH model.** (A) topological phases. (B) trivial phases. The gray lines represent theoretical values, and the red circles are experimental values.

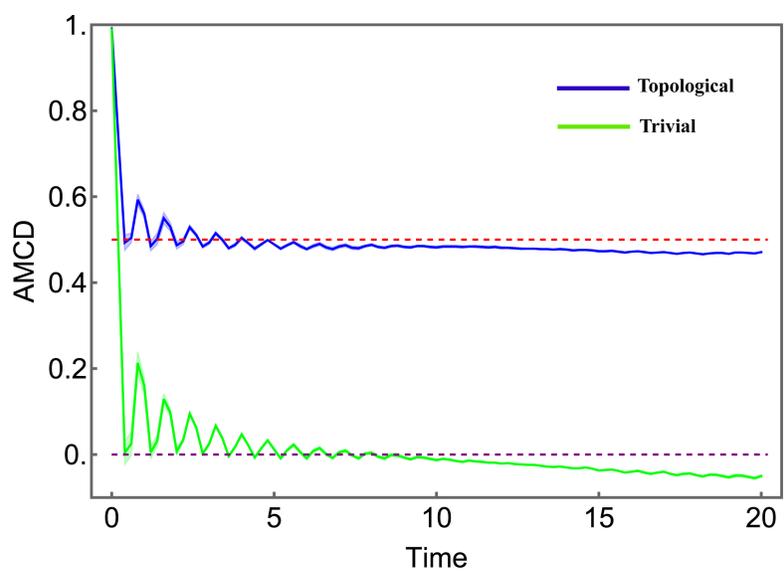

**Fig. S21 Long-time AMCDs on dimension x in topological and trivial phases of the 2D SSH model.**

**Table S1 Experimental details in the simulations of fast hitting dynamics on eRGTs, eCubes, eNets, and eGrids.** The evolution time and evolution steps are listed. At each time step, we measured the probability of the walker being at each vertex of the constructed graph, and then obtained the probability distribution of each vertex evolving with time. The average of the similarity between the experimental obtained and theoretical probability distributions of each vertex is also presented for each graph.

| | | | | | | |
|---|---|---|---|---|---|---|
| eRGT | Layer depth | 6 | 10 | - | | |
| | Evolution time | 10 | 15 | | | |
| | Evolution steps | 1000 | 1500 | | | |
| | Average fidelity | 99.59±0.18% | 96.28±1.86% | | | |
| eCube | Layer depth | 4 | 6 | 8 | - | |
| | Evolution time | 12 | 18 | 24 | | |
| | Evolution steps | 1200 | 1800 | 2400 | | |
| | Average fidelity | 99.49±0.27% | 99.22±0.42% | 97.98±1.06% | | |
| eNet | Layer depth | 4 | 6 | 8 | 10 | 12 |
| | Evolution time | 10 | 20 | 20 | 20 | 50 |
| | Evolution steps | 1000 | 1000 | 1000 | 1000 | 1000 |
| | Average fidelity | 99.77±0.12% | 99.59±0.17% | 99.22±0.24% | 98.61±0.42% | 97.20±0.92% |
| | Layer depth | 14 | 16 | 18 | 20 | - |
| | Evolution time | 50 | 50 | 100 | 100 | |
| | Evolution steps | 1000 | 1000 | 1000 | 1000 | |
| | Average fidelity | 98.73±0.31% | 98.36±0.40% | 97.64±0.67% | 96.92±0.77% | |
| eGrid | Layer depth | 4 | 8 | 12 | 16 | 20 |
| | Evolution time | 100 | 100 | 100 | 100 | 100 |
| | Evolution steps | 1000 | 1000 | 1000 | 1000 | 1000 |
| | Average fidelity | 99.82±0.14% | 99.72±0.12% | 99.42±0.23% | 98.89±0.42% | 97.99±0.74% |
| | Layer depth | 24 | 28 | 32 | 36 | - |
| | Evolution time | 100 | 100 | 100 | 100 | |
| | Evolution steps | 1000 | 1000 | 1000 | 1000 | |
| | Average fidelity | 96.36±1.49% | 98.22±0.60% | 97.73±0.94% | 96.41±1.55% | |

**Table S2 Comparisons of hitting on eRGTs and eCubes between the CTQWs and CTRWs.** The experimentally tested graphs are shown in Fig. S8 and Fig. S9. The theoretical predictions and experimental results of hitting time and optimal hitting efficiency are listed for each test case.

| Graph | Layer depth | Classical optimal hitting efficiency | Classical hitting time | Quantum optimal hitting efficiency | | Quantum hitting time | |
|---|---|---|---|---|---|---|---|
| | | | | Theoretical | Experimental | Theoretical | Experimental |
| eRGT | 6 | $4.76042 \times 10^{-2}$ | 5.120 | 0.887546 | 0.869505 | 1.812 | 1.80 |
| | 10 | $9.52067 \times 10^{-3}$ | 9.045 | 0.679780 | 0.705873 | 2.598 | 2.64 |
| | 14 | $2.15008 \times 10^{-3}$ | 13.000 | 0.596916 | - | 3.400 | - |
| | 18 | $5.11980 \times 10^{-4}$ | 17.325 | 0.521403 | - | 4.170 | - |
| | 22 | $1.24980 \times 10^{-4}$ | 21.600 | 0.463907 | - | 4.932 | - |
| eCube | 4 | $9.99744 \times 10^{-2}$ | 7.368 | 1.0 | 0.989880 | 3.048 | 3.11 |
| | 6 | $2.77778 \times 10^{-2}$ | 11.097 | 1.0 | 0.976987 | 4.590 | 4.67 |
| | 8 | $7.35294 \times 10^{-3}$ | 14.736 | 1.0 | 0.958245 | 6.144 | 6.36 |
| | 10 | $1.89394 \times 10^{-3}$ | 18.450 | 1.0 | - | 7.700 | - |
| | 12 | $4.80769 \times 10^{-4}$ | 22.248 | 1.0 | - | 9.288 | - |
| | 14 | $1.21124 \times 10^{-4}$ | 26.313 | 1.0 | - | 10.878 | - |

**Table S3 Comparisons of hitting on eNets and eGrids between the CTQWs and CTRWs.** The experimentally tested graphs are shown in Fig. S10 and Fig. S11. For each test case, the theoretical predictions and experimental results of hitting time and optimal hitting efficiency are listed.

| Graph | Layer depth | Classical optimal hitting efficiency | Classical hitting time | Quantum optimal hitting efficiency | | Quantum hitting time | |
|---|---|---|---|---|---|---|---|
| | | | | Theoretical | Experimental | Theoretical | Experimental |
| eNet | 4 | $1.00000 \times 10^{-2}$ | 3.696 | 0.999994 | 0.995572 | 1.572 | 1.58 |
| | 6 | $4.76190 \times 10^{-2}$ | 7.104 | 0.562500 | 0.539417 | 2.094 | 2.08 |
| | 8 | $2.77778 \times 10^{-2}$ | 11.808 | 0.408770 | 0.395388 | 2.656 | 2.66 |
| | 10 | $1.81817 \times 10^{-2}$ | 17.600 | 0.313299 | 0.302388 | 3.210 | 3.22 |
| | 12 | $1.28204 \times 10^{-2}$ | 24.816 | 0.251649 | 0.237820 | 3.756 | 3.80 |
| | 14 | $9.52364 \times 10^{-3}$ | 32.928 | 0.208645 | 0.185132 | 4.284 | 4.30 |
| | 16 | $7.35272 \times 10^{-3}$ | 42.432 | 0.177123 | 0.146451 | 4.816 | 4.90 |
| | 18 | $5.84770 \times 10^{-3}$ | 53.424 | 0.153156 | 0.116049 | 5.364 | 5.40 |
| | 20 | $4.76162 \times 10^{-3}$ | 64.800 | 0.134397 | 0.103111 | 5.880 | 5.80 |
| eGrid | 4 | $1.66667 \times 10^{-1}$ | 7.26 | 0.999996 | 0.985789 | 2.220 | 2.20 |
| | 8 | $6.66667 \times 10^{-2}$ | 18.30 | 0.888088 | 0.862709 | 3.380 | 3.40 |
| | 12 | $3.57143 \times 10^{-2}$ | 34.44 | 0.777954 | 0.775542 | 4.494 | 4.60 |
| | 16 | $2.22222 \times 10^{-2}$ | 55.35 | 0.685765 | 0.633813 | 5.580 | 5.50 |
| | 20 | $1.51515 \times 10^{-2}$ | 81.18 | 0.610303 | 0.578561 | 6.655 | 6.60 |
| | 24 | $1.09890 \times 10^{-2}$ | 111.93 | 0.548198 | 0.539548 | 7.722 | 7.70 |
| | 28 | $8.33333 \times 10^{-3}$ | 147.60 | 0.496518 | 0.484345 | 8.775 | 8.70 |
| | 32 | $6.53595 \times 10^{-3}$ | 188.19 | 0.452991 | 0.455865 | 9.826 | 9.80 |
| | 36 | $5.26316 \times 10^{-3}$ | 233.70 | 0.415892 | 0.427314 | 10.868 | 10.90 |

**Table S4 Analyses of the optimal efficiency of quantum and classical hitting on the eRGTs and eCubes as a function of layer depth.** The linear, quadratic and exponential models are employed for the fitting of reciprocal of classical optimal hitting efficiency, while for quantum scenarios, polynomial models with different orders are used. The goodness of fitting is characterized by the coefficient of determination ($R^2$). For classical hitting, the exponential models accurately depict the decrease trend, in contrast to the polynomial decrease for the quantum hitting on eRGTs and constant optimal efficiency for the quantum hitting on eCubes.

| Graph | Data | Fitted Model | $R^2$ |
|---|---|---|---|
| eRGT | Classical optimal hitting efficiency | $e_C = (445.22N - 4123.95)^{-1}$ | 0.692146 |
| | | $e_C = (58.2867N^2 - 1186.81N + 5435.08)^{-1}$ | 0.957873 |
| | | $e_C = (2^{0.38291N + 1.08716N^{0.5} - 0.563504})^{-1}$ | 0.999987 |
| | Quantum optimal hitting efficiency | $e_Q = (1.91653N^{0.25} - 1.94763)^{-1}$ | 0.980006 |
| | | $e_Q = (1.11791N^{0.5} - 2.46N^{0.25} + 2.25954)^{-1}$ | 0.997479 |
| | | $e_Q = (0.248556N - 4.61536N^{0.5} + 12.2945N^{0.25} - 8.29801)^{-1}$ | 0.999566 |
| eCube | Classical optimal hitting efficiency | $e_C = (682.2N - 4298.8)^{-1}$ | 0.621109 |
| | | $e_C = (163.205N^2 - 2255.5N + 7016.78)^{-1}$ | 0.924452 |
| | | $e_C = (2^{1.11302N - 0.823482N^{0.5} + 0.514277})^{-1}$ | 0.999998 |
| | Quantum optimal hitting efficiency | $e_Q = 1.0$ | 1.0 |

**Table S5 Analyses of the quantum and classical hitting time on the eRGTs and eCubes as a function of layer depth.** The polynomial models with orders 0.5, 1.0, and 2.0 are employed for the fitting of classical and quantum hitting time. The goodness of fitting is characterized by the coefficient of determination ($R^2$). The linear models accurately depict the increase trend of classical and quantum hitting time.

| Graph | Data | Fitted Model | $R^2$ |
|---|---|---|---|
| eRGT | Classical hitting time | $t_C = 7.33494N^{0.5} - 13.608$ | 0.986749 |
| | | $t_C = 1.27126N - 1.72781N^{0.5} + 1.73945$ | 0.999943 |
| | | $t_C = 0.00346075N^2 + 1.00147N - 0.484519N^{0.5} + 0.180343$ | 0.999949 |
| | Quantum hitting time | $t_Q = 1.5337N^{0.5} - 2.17695$ | 0.990153 |
| | | $t_Q = 0.172048N + 0.165013N^{0.5} + 0.369596$ | 0.999983 |
| | | $t_Q = -0.000387963N^2 + 0.209587N - 0.0248452N^{0.5} + 0.627263$ | 0.999991 |
| eCube | Classical hitting time | $t_C = 10.8124N^{0.5} - 15.097$ | 0.989182 |
| | | $t_C = 2.1897N - 1.76954N^{0.5} + 2.19896$ | 0.999878 |
| | | $t_C = 0.0442811N^2 - 0.0366616N + 6.50871N^{0.5} - 6.21349$ | 0.999996 |
| | Quantum hitting time | $t_Q = 4.49581N^{0.5} - 6.28081$ | 0.990956 |
| | | $t_Q = 0.836313N - 0.309606N^{0.5} + 0.325068$ | 0.999997 |
| | | $t_Q = 0.00255098N^2 + 0.708055N + 0.167293N^{0.5} - 0.159562$ | 0.999999 |

**Table S6 Analyses of the optimal efficiency of quantum and classical hitting on the eNets and eGrids as a function of layer depth.** The polynomial models with orders up to 3 are employed for the fitting of reciprocal of classical optimal hitting efficiency, while for quantum scenarios, polynomial models with orders up to 2 are used. The goodness of fitting is characterized by the coefficient of determination ($R^2$). For classical hitting on eNets and eGrids, the quadratic models accurately depict the decrease trend, in contrast to the linear decrease for the quantum optimal hitting efficiency.

| Graph | Data | Fitted Model | $R^2$ |
|---|---|---|---|
| eNet | Classical optimal hitting efficiency | $e_C = (22.5057N - 54.7873)^{-1}$ | 0.95507 |
| | | $e_C = (0.500292N^2 + 0.492804N + 0.0371118)^{-1}$ | 1.0 |
| | | $e_C = (6.51133 \times 10^{-6} N^3 + 0.499862N^2 + 0.500857N - 0.00139357)^{-1}$ | 1.0 |
| | Quantum optimal hitting efficiency | $e_Q = (0.957997N^{0.5} - 9.18079)^{-1}$ | 0.957997 |
| | | $e_Q = (0.692132N - 1.94147N^{0.5} + 2.31099)^{-1}$ | 0.999767 |
| | | $e_Q = (0.00256265N^2 + 0.408089N - 0.448303N^{0.5} + 0.269687)^{-1}$ | 0.999983 |
| eGrid | Classical optimal hitting efficiency | $e_C = (10.75N - 139.0)^{-1}$ | 0.951034 |
| | | $e_C = (0.125N^2 + 0.75N + 1.0)^{-1}$ | 1.0 |
| | | $e_C = (-7.61582 \times 10^{-15} N^3 + 0.125N^2 + 0.75N + 1.0)^{-1}$ | 1.0 |
| | Quantum optimal hitting efficiency | $e_Q = (0.548556N^{0.5} - 0.648332)^{-1}$ | 0.955102 |
| | | $e_Q = (0.0630879N - 0.154771N^{0.5} + 1.06247)^{-1}$ | 0.999979 |
| | | $e_Q = (0.0000321314N^2 + 0.0570223N - 0.114693N^{0.5} + 0.996974)^{-1}$ | 0.999997 |

**Table S7 Analyses of the quantum and classical hitting time on the eNets and eGrids as a function of layer depth.** The polynomial models with orders up to 3 are employed for the fitting of classical hitting time, while for quantum scenarios, polynomial models with orders up to 2 are used. The goodness of fitting is characterized by the coefficient of determination ($R^2$). For classical hitting on eNets and eGrids, the quadratic models accurately depict the increase trend, in contrast to the linear increase for the quantum hitting time.

| Graph | Data | Fitted Model | $R^2$ |
|---|---|---|---|
| eNet | Classical hitting time | $t_C = 6.88248N - 54.7873$ | 0.955984 |
| | | $t_C = 0.151281N^2 + 0.226127N + 0.278945$ | 0.999954 |
| | | $t_C = -0.0000219533N^3 + 0.15273N^2 + 0.198975N + 0.408768$ | 0.999954 |
| | Quantum hitting time | $t_Q = 2.26786N^{0.5} - 3.85455$ | 0.982944 |
| | | $t_Q = 0.249004N + 0.139104N^{0.5} + 0.27978$ | 0.999991 |
| | | $t_Q = -0.0000583651N^2 + 0.255473N + 0.105097N^{0.5} + 0.326271$ | 0.999991 |
| eGrid | Classical hitting time | $t_C = 13.1154N - 168.599$ | 0.952273 |
| | | $t_C = 0.150453N^2 + 1.07921N - 0.0917544$ | 0.999992 |
| | | $t_C = -0.0000131668N^3 + 0.152033N^2 + 1.02735N + 0.297562$ | 0.999992 |
| | Quantum hitting time | $t_Q = 2.92867N^{0.5} - 5.77887$ | 0.975991 |
| | | $t_Q = 0.243702N + 0.211789N^{0.5} + 0.829773$ | 0.999998 |
| | | $t_Q = 0.0000260883N^2 + 0.238777N + 0.244329N^{0.5} + 0.776596$ | 0.999999 |

**Table S8 Experimental details in in the simulations of mixing dynamics on eNets and eGrids.** The evolution time and evolution steps are listed. At each time step, we measured the probability of the walker being at each vertex of the constructed graph, and then obtained the probability distribution of each vertex evolving with time. The average of the similarity between the experimental obtained and theoretical probability distributions of each vertex is also presented for each graph.

| | Layer depth | 2 | 3 | 4 | 5 | 6 |
|---|---|---|---|---|---|---|
| | Evolution time | 80 | 120 | 160 | 200 | 240 |
| | Evolution steps | 160 | 240 | 320 | 400 | 480 |
| eNet | Average fidelity | 99.73±0.13% | 99.56±0.18% | 99.23±0.25% | 98.70±0.42% | 97.87±0.72% |
| | Layer depth | 7 | 8 | 9 | 10 | |
| | Evolution time | 280 | 320 | 360 | 400 | - |
| | Evolution steps | 560 | 640 | 720 | 800 | |
| | Average fidelity | 96.55±1.10% | 98.29±0.42% | 97.36±0.70% | 96.35±0.85% | |
| | Layer depth | 2 | 4 | 6 | 8 | 10 |
| | Evolution time | 60 | 100 | 140 | 180 | 220 |
| | Evolution steps | 120 | 200 | 280 | 360 | 440 |
| eGrid | Average fidelity | 99.85±0.10% | 99.76±0.10% | 99.41±0.20% | 98.99±0.33% | 98.54±0.51% |
| | Layer depth | 12 | 14 | 16 | 18 | - |
| | Evolution time | 260 | 300 | 340 | 380 | |
| | Evolution steps | 520 | 600 | 680 | 760 | |
| | Average fidelity | 97.34±0.91% | 95.93±1.18% | 97.88±0.61% | 96.79±0.87% | |

**Table S9 Comparisons of mixing on eNets and eGrids between the CTQWs and CTRWs.** The experimentally tested graphs are shown in Fig. S10 and Fig. S11. For each test case, the theoretical prediction and experimental data of mixing time are listed.

| Graph | Layer depth | Classical mixing time | Quantum mixing time | |
|---|---|---|---|---|
| | | | Theoretical | Experimental |
| eNet | 4 | 2.92 | 4.24 | 4.5 |
| | 6 | 5.64 | 7.92 | 8.0 |
| | 8 | 9.44 | 10.24 | 10.5 |
| | 10 | 14.30 | 14.30 | 15.0 |
| | 12 | 19.92 | 15.96 | 16.0 |
| | 14 | 26.74 | 19.74 | 20.0 |
| | 16 | 34.72 | 22.24 | 22.5 |
| | 18 | 43.38 | 25.02 | 26.5 |
| | 20 | 53.20 | 27.60 | 28.5 |
| eGrid | 4 | 1.41 | 2.01 | 2.0 |
| | 8 | 3.50 | 4.10 | 4.0 |
| | 12 | 6.86 | 7.07 | 7.0 |
| | 16 | 10.98 | 9.36 | 9.5 |
| | 20 | 16.39 | 12.32 | 12.5 |
| | 24 | 22.75 | 14.69 | 14.5 |
| | 28 | 30.00 | 18.00 | 17.5 |
| | 32 | 38.25 | 21.42 | 20.5 |
| | 36 | 47.69 | 26.03 | 24.5 |

**Table S10 Analyses of the quantum and classical mixing time on the eNets and eGrids as a function of layer depth.** The polynomial models with orders up to 3 are employed for the fitting of classical mixing time, while for quantum scenarios, polynomial models with orders up to 2 are used. The goodness of fitting is characterized by the coefficient of determination ($R^2$). For classical mixing on eNets and eGrids, the quadratic models accurately depict the increase trend, in contrast to the linear increase for the quantum mixing time.

| Graph | Data | Fitted Model | $R^2$ |
|---|---|---|---|
| eNet | Classical mixing time | $t_C = 5.65407N - 45.318$ | 0.955271 |
| | | $t_C = 0.125392N^2 + 0.136815N + 0.32478$ | 1.0 |
| | | $t_C = -0.0000142485N^3 + 0.126333N^2 + 0.119193N + 0.40904$ | 1.0 |
| | Quantum mixing time | $t_Q = 21.5936N^{0.5} - 64.3754$ | 0.951169 |
| | | $t_Q = 1.58911N - 0.84614N^{0.5} + 0.0808777$ | 0.999932 |
| | | $t_Q = -0.000374979N^2 + 1.69978N - 1.68023N^{0.5} + 1.62323$ | 0.999944 |
| eGrid | Classical mixing time | $t_C = 2.70146N - 35.1098$ | 0.951116 |
| | | $t_C = 0.0313842N^2 + 0.190723N + 0.0404334$ | 0.999998 |
| | | $t_C = -7.20733 \times 10^{-8} N^3 + 0.0313928N^2 + 0.19044N + 0.0425645$ | 0.999998 |
| | Quantum mixing time | $t_Q = 14.7267N^{0.5} - 58.328$ | 0.942302 |
| | | $t_Q = 0.783933N - 0.289843N^{0.5} - 1.48779$ | 0.999908 |
| | | $t_Q = -0.000071368N^2 + 0.822986N - 0.674379N^{0.5} - 0.598094$ | 0.999916 |

**Table S11 More experimental details in centrality measure on eScale-free networks.** The evolution time and evolution steps are listed. At each time step, we measured the probability of the walker being at each vertex of the network, and then obtained the probability distribution of each vertex evolving with time. The average of the fidelity between the experimental obtained and theoretical probability distributions of each vertex is also presented for each graph. The similarity (Similarity$_1$) between the experimental obtained CTQW centrality and theoretical CTQW centrality, and the similarity (Similarity$_2$) between the experimental obtained CTQW centrality and theoretical eigenvector centrality are also listed.

| Network | Evolution time | Evolution steps | Average fidelity | Similarity$_1$ | Similarity$_2$ |
|---|---|---|---|---|---|
| Fig. 3(A1) | 6.82 | 67 | 97.63±0.93% | 96.94% | 95.68% |
| Fig. S9 A | 6.73 | 62 | 97.58±0.91% | 94.92% | 95.09% |
| Fig. S9 D | 7.57 | 69 | 97.68±0.77% | 96.18% | 95.01% |

**Table S12 Experimental details of benchmark test of CTQW-based search over networks.** For each network size, we randomly generated 100 test cases. The evolution time is listed. For each time step of a test case, we measured the probability of the walker being at each vertex of the network, and then obtained the probability distribution of each vertex evolving with time. The average of the fidelity between the experimental obtained and theoretical probability distributions of each vertex is also presented for all test cases. After calculating the average of optimal search time and optimal search efficiency for test cases of each network size, we also compared the theoretical prediction with experimental results.

| Network size | Number of test cases | Evolution time | Average fidelity | Average optimal search time | | Average optimal search efficiency | |
|---|---|---|---|---|---|---|---|
| | | | | Simulation | Experiment | Simulation | Experiment |
| 36 | 100 | 15 | 97.52±0.36% | 7.15±2.64 | 7.14±2.81 | 71.70±14.53% | 67.83±14.46% |
| 55 | 100 | 20 | 97.88±0.26% | 8.69±1.98 | 8.56±2.14 | 60.71±14.78% | 58.77±13.04% |
| 78 | 100 | 22 | 97.55±0.25% | 10.80±2.62 | 10.91±2.54 | 61.00±13.52% | 60.10±11.61% |
| 105 | 100 | 24 | 96.94±0.41% | 11.30±2.39 | 11.12±3.75 | 60.35±14.16% | 58.76±12.94% |
| 120 | 100 | 24 | 95.99±0.31% | 12.25±1.77 | 12.46±2.75 | 62.40±14.35% | 58.54±12.82% |
| 136 | 100 | 25 | 96.66±0.37% | 14.20±2.68 | 14.00±2.45 | 70.42±12.75% | 66.00±11.83% |
| 153 | 100 | 25 | 96.34±0.33% | 14.42±2.10 | 14.33±2.53 | 70.50±10.47% | 65.68±9.32% |
| 171 | 100 | 25 | 96.04±0.35% | 14.48±2.31 | 14.51±2.72 | 70.27±11.99% | 65.98±10.50% |
| 190 | 100 | 25 | 95.95±0.34% | 14.93±2.02 | 14.74±2.52 | 71.97±9.95% | 66.42±9.81% |
| 210 | 100 | 25 | 95.46±0.41% | 15.05±1.84 | 15.15±1.78 | 71.75±11.30% | 66.48±9.41% |

**Table S13 Total variance distance (TVD) of graph certificates of isomorphic and non-isomorphic graph pairs.** The theoretical predictions and experimentally obtained results are both listed.

| Evolution step | Isomorphic TVD | | Non-isomorphic TVD | |
|---|---|---|---|---|
| | Simulation | Experiment | Simulation | Experiment |
| 1 | 0 | 0.1077 | 3.0169 | 2.4954 |
| 2 | 0 | 0.1493 | 0.7460 | 0.5284 |
| 3 | 0 | 0.0911 | 0.5874 | 0.5305 |
| 4 | 0 | 0.0678 | 1.1805 | 1.0529 |
| 5 | 0 | 0.0944 | 0.3322 | 0.2881 |
| 6 | 0 | 0.0845 | 1.8110 | 1.5579 |
| 7 | 0 | 0.1033 | 1.7735 | 1.1834 |
| 8 | 0 | 0.0980 | 0.4268 | 0.4113 |
| 9 | 0 | 0.0864 | 1.4083 | 1.1567 |
| 10 | 0 | 0.1308 | 1.1936 | 0.9177 |
| Average | 0 | 0.1013 | 1.2476 | 1.0122 |

**Table S14 Experimental details of simulating the topological phases of 2D SSH and BBH model.** For each model, we both simulate its bulk topology in trivial and topological phases. The evolution time and evolution steps are listed. When simulating the bulk topology, for each time step, we measured the probability distributions of single particle or two particles on vertices in the related model and then calculated the averaged values of corresponding bulk observables up to this time step. The averaged fidelities between the experimental and theoretical probability distributions of each model in different phases are also presented. We also obtained the asymptotic values of long-time averaged values of extended MCDs and MCQMs with pretty high precision.

| Model | Evolution time | Time step | Averaged fidelity | | $\langle \vec{m} \rangle_{aver}$ (or $\langle m_q \rangle_{aver}$) | |
|---|---|---|---|---|---|---|
| | | | Trivial | Nontrivial | Trivial | Nontrivial |
| 2D SSH | $t = 20$ | 0.04 | 97.43±0.50% | 96.72±0.78% | (-0.046±0.004, 0.013±0.003) | (0.470±0.002, 0.498±0.001) |
| BBH | $t = 20$ | 0.04 | 96.98±-.73% | 97.66±0.77% | 0.014±0.003 | 0.513±0.002 |


\fi

\end{document}